\documentclass[11pt]{article}

\usepackage{graphicx}
\usepackage{float}
\usepackage{mathtools}
\usepackage{scalerel}
\usepackage{amssymb}
\usepackage{amsfonts}
\usepackage{dsfont}
\usepackage{mathrsfs}
\usepackage{enumitem}
\usepackage{amsthm}
\usepackage{multirow}
\usepackage{subcaption}
\usepackage{chet}
\usepackage[font=small,format=hang,labelfont={sf,bf}]{caption}

\usepackage{arydshln}

\newcommand{\lsp}{\hspace{1pt}}

\newcommand{\lnsp}{\hspace{-1pt}}

\newcommand{\veps}{\varepsilon}

\renewcommand{\geq}{\geqslant}

\allowdisplaybreaks

\definecolor{darkblue}{rgb}{0.1,0.1,0.7}
\hypersetup{colorlinks,
           linkcolor={darkblue},
           citecolor={darkblue},
           urlcolor={darkblue}
}

\newcommand{\1}{{\mathds{1}}}
\renewcommand{\O}{{\mathcal O}}

\newcommand{\MN}{M\hspace{-0.25pt}N\xspace}
\newcommand{\MNmath}{\ensuremath{M\lnsp N}}

\preprint{LA-UR-21-20310}

\date{January 2021}

\title{Perturbative and Nonperturbative Studies\\[6pt]
of CFTs with \MN Global Symmetry}

\author{Johan Henriksson$^{a,b,c}$ and Andreas Stergiou$^{d}$}

\affiliation{$^a$Dipartimento di Fisica E.\ Fermi, Universit\`a di Pisa, \emph{and} INFN, Sezione di Pisa,\\\vspace{-3pt} Largo Bruno Pontecorvo 3, 56127 Pisa, Italy\\
$^b$Lincoln College, University of Oxford, Turl Street, Oxford, OX1 3DR, UK
\\
$^c$Mathematical Institute, University of Oxford, Woodstock Road, Oxford, OX2 6GG, UK
\\
$^d$Theoretical Division, MS B285, Los Alamos National Laboratory, Los
Alamos, NM 87545, USA}

\abstract{Fixed points in three dimensions described by conformal field
theories with $\MNmath_{m,n}= O(m)^n\rtimes S_n$ global symmetry have
extensive applications in critical phenomena.  Associated experimental data
for $m=n=2$ suggest the existence of two non-trivial fixed points, while
the $\varepsilon$ expansion predicts only one, resulting in a puzzling state
of affairs. A recent numerical conformal bootstrap study has found two
kinks for small values of the parameters $m$ and $n$, with critical
exponents in good agreement with experimental determinations in the $m=n=2$
case. In this paper we investigate the fate of the corresponding fixed
points as we vary the parameters $m$ and $n$. We find that one family of
kinks approaches a perturbative limit as $m$ increases, and using large spin perturbation theory we construct a large $m$ expansion that fits well with
the numerical data. This new expansion, akin to the large $N$ expansion of
critical $O(N)$ models, is compatible with the fixed point found in the
$\varepsilon$ expansion. For the other family of kinks, we find that it
persists only for $n=2$, where for large $m$ it approaches a
non-perturbative limit with $\Delta_\phi\approx 0.75$. We investigate the
spectrum in the case $\MNmath_{100,2}$ and find consistency with
expectations from the lightcone bootstrap.
}

\begin{document}

\maketitle

\toc
\newpage

\newsec{Introduction}

Second-order phase transitions display scale-invariant physics and are
widely believed to be described by conformal field theories (CFTs), which
arise at fixed points of the renormalization group (RG) flow. Due to
universality, the physics at these phase transitions is independent of the
underlying microscopic degrees of freedom, which means that the same CFT
may describe a variety of systems. Many important applications of
three-dimensional CFTs arise for non-zero temperature phase transitions,
for instance critical liquid-vapor transitions, transitions between
magnetic phases and structural phase transitions.

The observables of a conformal field theory, such as the critical
exponents, can be extracted from the CFT data, which are the scaling
dimensions and structure constants (OPE coefficients) of the local
operators in the theory. One principal goal in the theory of critical
phenomena is therefore to make precise determinations of the CFT data. A
useful tool in studying CFTs relevant for three-dimensional systems is the
Landau--Ginzburg--Wilson description, in which one writes down a quantum
field theory with quartic interactions preserving a given global symmetry
group. By tuning mass parameters (equivalent to tuning the temperature in
experiments), the field theory flows under the RG to a fixed point
preserving the same (or larger) global symmetry. Methods within this
paradigm, such as the $\varepsilon$ expansion \cite{Wilson:1971dc}, produce
in many cases values of the critical exponents that match well with
experiments; see \cite{Pelissetto:2000ek} for an extensive review.

Despite the remarkable success of the mentioned paradigm for
many systems, including those with emergent $O(N)$ symmetry, one cannot
rule out the existence of additional CFTs, not captured by the
Landau--Ginzburg--Wilson description but still relevant for experimental
realizations.  An interesting case is systems with global symmetry group
$\MNmath_{m,n}=O(m)^n\rtimes S_n$. The most general Lagrangian that
preserves $\MNmath_{m,n}$ symmetry is
\eqn{\mathscr{L}=\tfrac12\lsp\partial_\mu\phi_i\lsp\partial^\mu\phi_i
+\tfrac18\lsp\lambda\lsp(\phi^2)^2 +\tfrac{1}{24}\lsp g\lsp
[(\phi_1^2+\cdots+\phi_m^2)^2
+\cdots+(\phi_{m(n-1)+1}^2+\cdots+\phi_{mn}^2)^2]\,,}[LagMN]
where $\phi_i$ is an $mn$-dimensional vector under $\MNmath_{m,n}$,
$\phi^2=\phi_i\phi_i$, and we keep only the kinetic term and the quartic
interaction terms. For the two-coupling theory \eqref{LagMN}, the
$\varepsilon$ expansion predicts only one fully-interacting fixed point
with this global symmetry.\footnote{In addition to this fixed point, one
finds also the free theory with $g=\lambda=0$, the $O(mn)$ symmetric theory
with $g=0$ and $n$ decoupled $O(m)$ models with $\lambda=0$.} For the
experimentally accessible case of $m=n=2$, when the $O(2)^2\rtimes S_2$
theory in \LagMN is equivalent to the more commonly discussed
$O(2)^2/\mathbb{Z}_2$ theory~\cite{Osborn:2017ucf}, the critical exponents
derived from this fixed point have not been successful in matching those
measured in experiments with helimagnets and XY stacked triangular
antiferromagnets, which cluster in two distinct regions; see
Table~\ref{tab:experiments}.
\begin{table}
\centering
\caption{Experimental results for phase transitions described by
$\MNmath_{2,2}$ theory. These values are compiled from
\cite{Pelissetto:2000ek, Delamotte:2003Dw} and references therein.}\label{tab:experiments}
\begin{tabular}{|l|cc|}
\hline
{$\boldsymbol{\MNmath_{2,2}}$}& $\beta$ &$\nu$
\\\hline
XY STAs & $0.24(2)$& $0.55(5)$
\\
Tb & $0.23(4)$& $0.53(4)$
\\\hline
Ho, Dy & $0.39(4)$ & $0.57(4)$
\\
NbO$_2$ & $0.40^{+0.04}_{-0.07}$ &
\\\hline
\end{tabular}
\end{table}
Additionally, the fixed point in the $\varepsilon$ expansion appears to have $g<0$, which is inconsistent
with the expected chiral universality class that should describe phase
transitions in these systems~\cite{KawamuraIII, Pelissetto:2000ek}.\foot{In
much of the literature, e.g.\ \cite{KawamuraIII, Pelissetto:2000ek}, a
coupling $v$ is used instead of our $g$ in \LagMN. The coupling $v$ and our
coupling $g$ have opposite signs, so the chiral region, defined by $v>0$ in
the literature, corresponds to $g<0$ in our notation.} Evidence for the
existence of further non-perturbative fixed points in the chiral region has
been offered~\cite{Pelissetto:2000ne, DePrato:2003ra,
Calabrese:2004nt}, but this has been disputed by other
authors~\cite{Tissier:2000tz, Tissier:2001uk, Delamotte:2003Dw,
Delamotte:2016acs, Delamotte_2008, Delamotte:2010ba, Delamotte_2010}.

This contradictory set of observations motivated the recent study of \MN symmetric theories
using the non-perturbative (numerical) conformal bootstrap \cite{Stergiou:2019dcv}.
This method, proposed in \cite{Rattazzi:2008pe} and extensively reviewed in
\cite{Poland:2018epd}, makes no assumptions on the underlying microscopic
description and studies CFTs based only on global and conformal symmetry,
unitarity, and consistency with the operator algebra (crossing symmetry).
\begin{figure}[ht]
\centering
\includegraphics{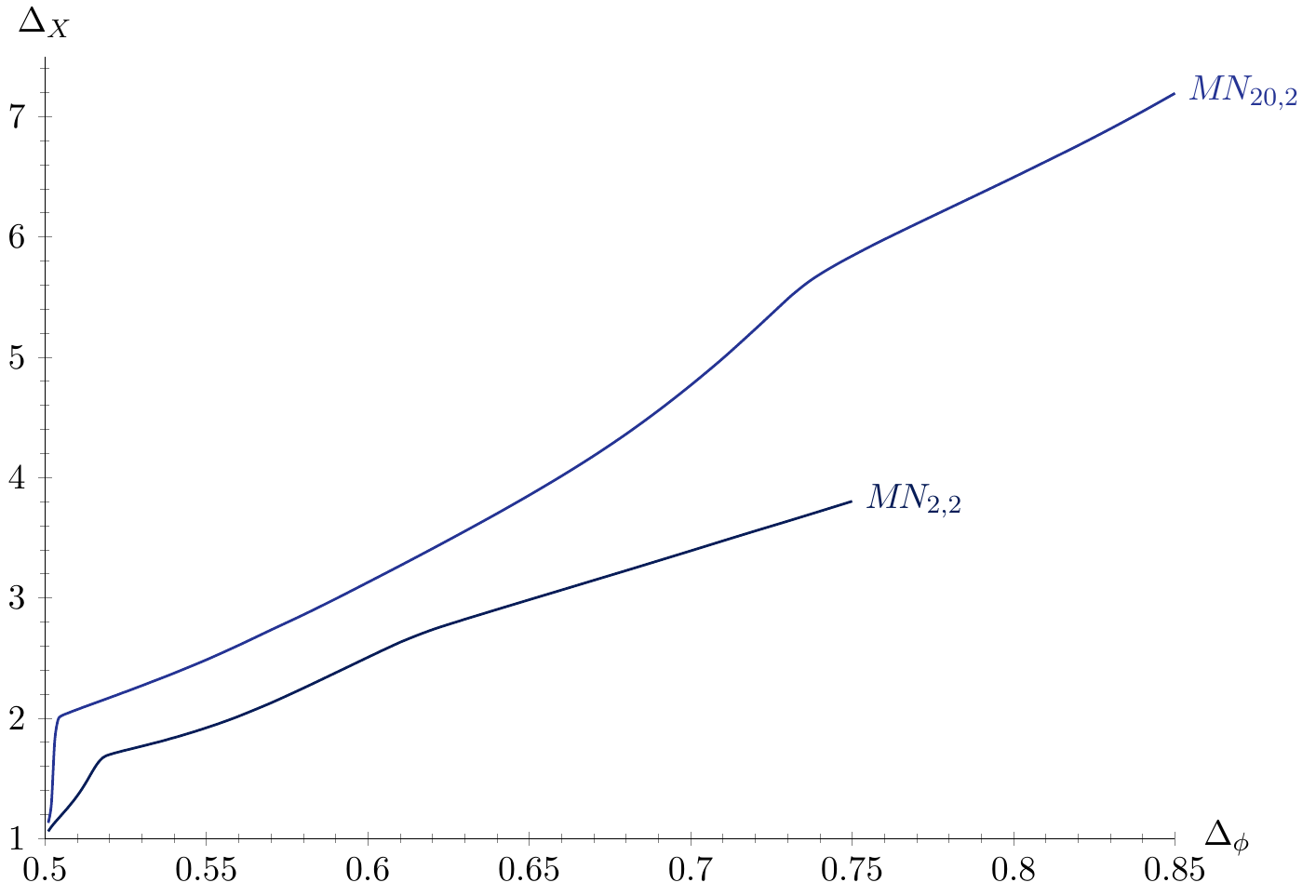}
\caption{Bounds in the $(\Delta_\phi,\Delta_X)$ plane for 3D CFTs with
$\MNmath_{m,n}$ global symmetry with $(m,n)=(2,2)$ and $(m,n)=(20,2)$.
The allowed region lies below the curves for the corresponding parameter values. These bounds are obtained with the use of \texttt{PyCFTBoot}
  \cite{Behan:2016dtz} with parameters $\texttt{n\_max=9}$,
$\texttt{m\_max=6}$, $\texttt{k\_max=36}$ and $\texttt{l\_max=26}$.}
\label{fig:DeltaX-2-2-20-2}
\end{figure}
In agreement with the experimental data, the conformal bootstrap study in
\cite{Stergiou:2019dcv} found evidence for the existence of two distinct
CFTs with $\MNmath_{2,2}$ symmetry, as can be seen from
Fig.~\ref{fig:DeltaX-2-2-20-2}. This figure displays bounds on operator
dimensions in the $(\Delta_\phi,\Delta_X)$ plane, where $\phi$ and $X$
respectively denote the smallest dimension operators transforming in the
vector and a certain rank-two representation of the \MN symmetry group. The
region below the curves is the allowed parameter space in the respective
theories. In various applications of the conformal bootstrap, it has been observed that a kink in the boundary of the allowed
region is related to the existence of a conformal field theory with parameters
near the location of the kink, and from the two kinks of the
$\MNmath_{2,2}$ curve in Fig.~\ref{fig:DeltaX-2-2-20-2} the following
values for the critical exponents were derived \cite{Stergiou:2019dcv}:
\begin{equation}
  \text{kink 1: }\beta=0.293(3)\,,\ \nu=0.566(6)\,,\qquad \text{kink 2:
  }\beta=0.355(5)\,,\ \nu=0.576(8)\,.
\end{equation}
While the critical exponents corresponding to the second kink show
reasonable agreement, within uncertainties, with the results
of~\cite{Pelissetto:2001fi, Calabrese:2004nt}, neither set of values is
compatible with the predictions from the $\varepsilon$ expansion, which
gives $\beta=0.370(5)$ and $\nu=0.715(10)$
\cite{Mudrov:2001yr,Mudrov2}.\foot{The corresponding fixed point is called
``complex cubic'' in \cite{Mudrov:2001yr,Mudrov2}. Regarding RG stability
of this fixed point relative to the decoupled $O(2)$ one, we note that
since the scaling dimension of the first singlet of the $O(2)$ model is
slightly above $1.5$~\cite{Chester:2019ifh}, this means that the decoupled
$O(2)$ theory is stable in $d=3$ \cite{Aharony1973}.} Although it was speculated in
\cite{Stergiou:2019dcv} that the first kink may be related to the
$\varepsilon$ expansion through the large $m$ limit, the results of that
study were not sufficient to make any conclusive statements.

The purpose of the present paper is to make a more systematic study of the
hypothetical CFTs living at the two kinks, by investigating the behavior of
numerical bounds for various $(m,n)$. More specifically, we perform
numerical conformal bootstrap studies for varying values of $m$, keeping
$n$ fixed (for most of the work we keep $n=2$, but we also
obtain a bound for $m=100$ and $n=3,4$).  For $n=2$ we find that, as we
increase $m$, both kinks continue to exist, and for large $m$ the bound in
the $(\Delta_\phi,\Delta_X)$ plane attain a profile similar to the case
$\MNmath_{20,2}$ displayed in Fig.~\ref{fig:DeltaX-2-2-20-2}. The large $m$
behavior of the CFT at each kink is subsequently studied.

First, we focus on the first kink, which approaches the value
$(\Delta_\phi,\Delta_X)=(0.5,\, 2)$ as $m\to\infty$. In
\cite{Stergiou:2019dcv}, it was observed that this limit is compatible with
the results in the $\varepsilon$ expansion \cite{Mukamel1975b,
Mudrov:2001rt, Mudrov2,Osborn:2017ucf} expanded at large $m$, where indeed
$\Delta_\phi\to \frac{d-2}2$ and $\Delta_X\to 2$ up to $m^{-1}$
corrections. We show here that the $m^{-1}$ corrections can be computed in
a perturbative expansion similar to the usual large $N$ expansion of the
$O(N)$ model (see e.g.\ \cite{Moshe:2003xn}), where now the operator $X$
acts as the Hubbard--Stratonovich auxiliary field. Specifically, we use the
analytic conformal bootstrap method of large spin perturbation theory, developed in \cite{Alday:2016njk,Alday:2016jfr,
Alday:2019clp, Henriksson:2020fqi}, to compute $m^{-1}$ corrections to
scaling dimensions and OPE coefficients. This expansion is valid for all
spacetime dimensions $d\in(2,4]$. Near $d=4$, the results agree with the
$\varepsilon$ expansion and in $d=3$ we get good agreement with the
non-perturbative bootstrap results for the first kink. This gives
substantial evidence that we should view the first kink as describing a
perturbative CFT with \MN symmetry, existing for a range of $(m,n)$ and $d$
and connected to the $\varepsilon$ expansion via the large $m$ limit.

Second, we study the second kink for $n=2$ and increasing $m$. The bounds
in Fig.~\ref{fig:DeltaX-2-2-20-2} reveal that, as we increase $m$, the
values of the scaling dimensions $\Delta_\phi$ and $\Delta_X$ corresponding
to this kink move far away from the free theory values. We perform a single
correlator numerical bootstrap study where we search for bounds in the
$(\Delta_\phi,\Delta_X)$ plane for increasing values of $m$, with the hope
of finding a limit point at infinite $m$ for the position of the kink. The
results show that the second kink continues to exist for all values of $m$
studied, and that the position in the $\Delta_\phi$ direction appears to
stabilize near the value $0.75$. The position in the $\Delta_X$ direction
takes a value $\Delta_X\gtrsim6$, but is highly sensitive to the numerical
precision of the computation (number of derivatives in the functional used
for the numerical bootstrap computations). These values for $\Delta_\phi$
and $\Delta_X$ show that, if there is a CFT corresponding to the second
kink, it must be of a non-perturbative type.

To investigate further the potential CFT corresponding to the second kink,
we focus on the case $\MNmath_{100,2}$ and increase the numerical
precision. We use the extremal functional method, developed in
\cite{ElShowk:2012hu}, to extract information about the spectrum of
operators in the $\phi\times\phi$ OPE. The results give some hints of an
organization of the leading twist operators in twist families, as must be
the case according to the lightcone bootstrap
\cite{Fitzpatrick:2012yx, Komargodski:2012ek, Simmons-Duffin:2016wlq}.  For
fixed $m=100$ we also study $\MNmath_{100,n}$ for $n=3 $ and $n=4$, but for
these values we find no second kink in the $(\Delta_\phi,\Delta_X)$ bound.

This paper is organized as follows. In section~\ref{sec:review} we explain
how to study \MN symmetric CFTs in the bootstrap approach using crossing
symmetry and unitarity. We introduce relevant notation and review the known
perturbative results of the $\varepsilon$ expansion. In section
\ref{sec:kink1} we show that the CFT corresponding to the first kink can be
matched with a perturbative large $m$ expansion, which we construct using
large spin perturbation theory. Further, we comment on the connection to the
$\varepsilon$ expansion. In section \ref{sec:kink2} we use the
non-perturbative numerical bootstrap to study the second kink, and discuss
the twist families of the spectrum for the representative case
$\MNmath_{100,2}$. We finish with a discussion, and include some explicit
results in an appendix.

For our numerical computations we have used
\texttt{PyCFTBoot}~\cite{Behan:2016dtz}, \texttt{qboot}~\cite{Go:2020ahx}
and \texttt{SDPB}~\cite{Landry:2019qug}.

\newsec{Review}\label{sec:review}

In this section we briefly review the constraints from unitarity and crossing symmetry on conformal field theories, with emphasis on theories with global \MN symmetry. We then summarize the results from previous studies in the $\varepsilon$ expansion.

\subsec{Unitarity and crossing in the presence of
\texorpdfstring{$\MNmath_{m,n}$}{MN\_m,n} symmetry}

We consider the four-point correlator of $\phi^{i}$, $i=1,\ldots,mn$,
transforming in the vector representation $V$ of the $\MNmath_{m,n}=O(m)^n
\rtimes S_n$ global symmetry. More precisely, $O(m)$ acts by rotating the
fields within each of the $n$ groups of $m$ fields, and $S_n$ permutes
these groups. In terms of the conformal cross-ratios
$u=\frac{x_{12}^2x_{34}^2}{x_{13}^2x_{24}^2}$ and
$v=\frac{x_{14}^2x_{23}^2}{x_{13}^2x_{24}^2}$, with $x_{ij}=|x_i-x_j|$, the
correlator takes the form
\begin{equation}\label{eq:correlator}
\langle \phi^i(x_1)\phi^j(x_2)\phi^k(x_3)\phi^l(x_4)\rangle =
\frac1{x_{12}^{2\Delta_\phi}x_{34}^{2\Delta_\phi}}
\sum_{R=S,X,Y,Z,A,B}\mathbf T_R^{ijkl} \mathcal G_R(u,v)\,,
\end{equation}
where $\mathbf T_R$ are projection tensors for the representations $R$ in
the tensor product $V\otimes V=S\oplus X\oplus Y\oplus Z\oplus A\oplus B$.
$S$ denotes the singlet representation with $\mathbf
T_S^{ijkl}=\frac1{mn}\delta^{ij}\delta^{kl}$, while the remaining symbols
denote rank-two symmetric ($X$, $Y$ and $Z$) and antisymmetric ($A$ and
$B$) representations. We will not need the precise form of these projection
tensors, which can be found in \cite{Stergiou:2019dcv}. Each function
$\mathcal G_R(u,v)$ admits a decomposition in conformal blocks,
\begin{equation}\label{eq:CBdecompR}
\mathcal G_R(u,v)=\sum_{\O\in R} (-1)^{\ell_\O}\lsp\lambda^2_{\O}\lsp
g_{\Delta_\O,\ell_\O}(u,v)\,,
\end{equation}
where the sum runs over conformal primary operators of dimension
$\Delta_\O$ and spin $\ell_\O$ transforming in the representation $R$. The
conformal blocks, denoted by $g_{\Delta,\ell}(u,v)$, sum up the
contribution to the correlator of a given primary and all its descendants,
and are functions depending only on the cross-ratios and the dimension
and spin of the primary.

We will adopt a notation where we denote by $R$, $R'$, $R''$ etc.\ the
smallest dimension scalars in the representation $R$, and likewise by
$R_\ell$, $R'_\ell$, $R''_\ell$ etc.\ the smallest dimension spin $\ell$
operators in the representation $R$. In any unitary CFT, the $S$
representation will contain the identity operator $\1$ with
$\Delta=\ell=0$,\footnote{This means we will denote by $S$ the smallest
dimension operator different from the identity.} and the stress-energy
tensor $T=S_2$ with $\Delta_T=d$. In the presence of a continuous
global symmetry, a CFT contains in addition a conserved Noether current
$J^{\mu}$ with $\Delta_J=d-1$, which in our case resides in the $A$
representation: $J=A_1$.

Unitarity imposes constraints on the decomposition \eqref{eq:CBdecompR}. Reality
of the three-point functions $\langle \phi(x_1)\phi(x_2)\O(x_3)\rangle$ implies
positivity of the expansion coefficients $\lambda^2_{\O}$, and positivity of
two-point functions of descendants implies unitarity bounds, namely
\begin{equation}\label{eq:UB}
  \Delta_R\geq\tfrac12(d-2)\,,\qquad \Delta_{R_\ell}\geq d-2+\ell\,,
\end{equation}
where the inequalities are saturated only for a free scalar and a conserved
current respectively. Moreover, a non-trivial consequence of unitarity is
Nachtmann's theorem \cite{Nachtmann:1973mr}, which states that the twists
of the leading singlet operators, $\tau_{S,\ell}=\Delta_{S_\ell}-\ell$,
form an upward convex function for all spin $\ell$ above some $\ell_0$, see
\cite{Kundu:2020gkz} for a recent discussion.

Crossing symmetry follows from the invariance of the correlator
\eqref{eq:correlator} under exchanging pairs of insertion points. The
invariance under $x_1\leftrightarrow x_2$ is satisfied by each conformal
block together with the fact that only operators of even/odd spins appear in each representation, while the invariance under $x_1\leftrightarrow x_3$ leads to a
non-trivial crossing equation which we will use. In the presence of global
$\MNmath_{m,n}$ symmetry, the crossing equation takes the form
\begin{equation}\label{eq:crossing}
\mathcal G_R(u,v)=\left(\frac uv\right)^{\Delta_\phi}\sum_{\widetilde R}
M_{R\widetilde R}\,\mathcal G_{\widetilde R}(v,u)\,,
\end{equation}
where the crossing matrix $M_{R\widetilde R}$ in the basis $\{S,X,Y,Z,A,B\}$ is given by
\begin{equation}
M_{RR'}=
\begin{pmatrix}
\frac{1}{m n} & \frac{1}{m n} & \frac{1}{m} & 1 & -1 & -1 \\
 \frac{n-1}{m n} & \frac{n-1}{m n} & \frac{n-1}{m} & -1 & 1-n & 1 \\
 \frac{(m-1) (m+2)}{2 m n} & \frac{(m-1) (m+2)}{2 m n} & \frac{m-2}{2 m} & 0 &
   \frac{m+2}{2} & 0 \\
 \frac{n-1}{2 n} & -\frac{1}{2 n} & 0 & \frac{1}{2} & 0 & \frac{1}{2} \\
 -\frac{m-1}{2 m n} & -\frac{m-1}{2 m n} & \frac{1}{2 m} & 0 & \frac{1}{2} & 0 \\
 -\frac{n-1}{2 n} & \frac{1}{2 n} & 0 & \frac{1}{2} & 0 & \frac{1}{2}
\end{pmatrix}.
\end{equation}
We refer to the left-hand side of \eqref{eq:crossing} as the direct channel, and to the right-hand side as the crossed channel. In the analytic bootstrap approach in section~\ref{sec:LSPT}, the crossing equation is expanded in the double lightcone limit $u\ll v\ll 1$, and the operators in the crossed channel will source corrections to the CFT data of the operators in the direct channel.

In the numerical bootstrap approach in
section~\ref{sec:numericalimplementation}, the crossing equation is
re-written in a form that treats the channels symmetrically.  This form is
given explicitly in \cite{Stergiou:2019dcv}, and the technical details can
be found in that paper.  The principles are the standard ones of the
numerical conformal bootstrap \cite{Rattazzi:2008pe,Poland:2018epd}: by
acting on the crossing equation with a family of functionals, positivity of
the squared OPE coefficients $\lambda^2_{\O}$ is turned into rigorous
inequalities which rule out large regions of the space of allowed operator
dimensions. For the theory at hand, we identify the potential fixed points
by observing kinks in the bound in the $(\Delta_\phi,\Delta_X)$ plane,
following \cite{Stergiou:2019dcv}. To gain more information about the CFT
at the position of the kink, we apply the extremal functional method
developed in \cite{ElShowk:2012hu}, which uses the fact that a functional
in the vicinity of the CFT should vanish when applied to the conformal
blocks of the operators present in the spectrum.

\subsec{Results from previous studies in the \texorpdfstring{$\varepsilon$}{epsilon} expansion}\label{sec:epsresults}

Conformal field theories with \MN symmetry have been studied in the $d=4-\varepsilon$ expansion over a long time, \cite{Mukamel1975b,Shpot1988,Shpot1989,Mudrov:2001rt,Mudrov:2001yr,Mudrov2}, most recently in section~5.2.2. of  \cite{Osborn:2017ucf}.
From the beta functions of the couplings $\lambda$ and $g$ in the Lagrangian \LagMN, four fixed points are found in the $\varepsilon$ expansion: $mn$ free fields, $n$
decoupled critical $O(m)$ models, the critical $O(mn)$ model, and the
perturbative \MN CFT.
We focus on the \MN CFT, which has $\lambda, g\ne0$. At this fixed point, the scaling dimensions of the leading scalar operators have an expansion of the form
\begin{align}\label{eq:deltaphi}
\Delta_\phi&=1-\frac\varepsilon2+ \frac{m (n-1) [(m + 2) m n - 10 m +
16]}{4\,C_{mn}^2}\varepsilon
^2+\gamma_\phi^{(3)}\varepsilon^3+O(\varepsilon^4)\,,
\\\label{eq:deltaS}
\Delta_S&=2-\varepsilon+\frac{6 m (n-1) }{C_{mn}}\varepsilon
+\gamma_S^{(2)}\varepsilon^2+\gamma_S^{(3)}\varepsilon^3+
+O(\varepsilon^4)\,,
\\\label{eq:deltaX}
\Delta_X&=2-\varepsilon+\frac{m   ((m+2) n-6)}{C_{mn}}\varepsilon
+\gamma_X^{(2)}\varepsilon^2+\gamma_X^{(3)}\varepsilon^3
+O(\varepsilon^4)\,,
\\\label{eq:deltaY}
\Delta_Y&=2-\varepsilon+\frac{2 m (n-1)}{C_{mn}}\varepsilon
+\gamma_Y^{(2)}\varepsilon^2+\gamma_Y^{(3)}\varepsilon^3
+O(\varepsilon^4)\,,
\\\label{eq:deltaZ}
\Delta_Z&=2-\varepsilon-\frac{2 (m-4)}{C_{mn}}\varepsilon
+\gamma_Z^{(2)}\varepsilon^2+\gamma_Z^{(3)}\varepsilon^3
+O(\varepsilon^4)\,,
\end{align}
where $C_{mn}=(m+8)mn-16(m-1)$. In the above expressions we have omitted
the explicit form of the order $\varepsilon^2$ and $\varepsilon^3$
corrections, which were derived in \cite{Osborn:2017ucf}\footnote{The
leading $m$ dependence is given in equation (5.104) and (5.105) in the
arXiv submission of \cite{Osborn:2017ucf}, where $N=mn$ and $\sigma=S$, $\rho_1=X$, $\rho_2=Z$ and $\rho_3=Y$.} and are available by an email request to the authors.
Moreover, the eigenvalues of the stability matrix are \cite{Osborn:2017ucf}
\begin{align}
\omega_1&=\varepsilon+ \omega_1^{(2)}\varepsilon^2+
\omega_1^{(3)}\varepsilon^3+ O(\varepsilon^4)\,,
\\
\omega_2&=-\frac{(m-4)(mn-4)}{C_{mn}}\varepsilon+
\omega_2^{(2)}\varepsilon^2+ \omega_2^{(3)}\varepsilon^3+
O(\varepsilon^4)\,,
\end{align}
which correspond to $\omega_i=\Delta_i-d$ for the singlet operators of $\phi^4$ type in the theory.

For the specific case of $m=2$, the order $\varepsilon^4$ renormalization was performed in \cite{Mudrov:2001yr,Mudrov2}, giving $\Delta_\phi$, $\Delta_S$, $\omega_1$ and $\omega_2$ to this order. For the physically relevant cases $n=2$ and $n=3$, a Borel--Leroy resummation was performed to give estimates for the critical exponents $\gamma$, $\nu$, $\eta$ in three dimensions; see section~\ref{sec:comparenum} below.

\newsec{The perturbative fixed point at large \texorpdfstring{$\boldsymbol{m}$}{m}}\label{sec:kink1}

In this section we will derive a large $m$ expansion for $\MNmath_{m,n}$ symmetric CFTs, and show that it gives predictions that match well with those found in the numerical bootstrap for the first kink. The existence of this expansion establishes the perturbative nature of the corresponding family of CFTs .

\subsec{Analytic expansion from large spin perturbation theory}\label{sec:LSPT}

Expanding the expressions \eqref{eq:deltaS}--\eqref{eq:deltaZ} for the
scalar operators at large $m$ we observe that
\begin{equation}
\Delta_X=2+O(m^{-1})\,,
\end{equation}
whereas the scalar operators in the $S$, $Y$ and $Z$ representations all
satisfy $\Delta=2-\varepsilon+O(m^{-1})$. This observation indicates that
there exists, for all $d\in(2,4]$, a large $m$ expansion where $\Delta_{
X}=2+O(m^{-1})$, and $\Delta_R=d-2+O(m^{-1})$ for $R=S,Y,Z$. These values
are consistent with a description in terms of Hubbard--Stratonovich
auxiliary fields, similar to the large $N$ expansion of the critical $O(N)$
model.

In \cite{Henriksson:2020fqi}, based on \cite{Alday:2016jfr,Alday:2019clp},
it was described how to use large spin perturbation theory to extract
properties of $\phi^4$ theories with Hubbard--Stratonovich auxiliary
fields. We will follow this approach, which means that we assume that at
large $m$ the operator spectrum is that of mean field theory for $\phi$
with $\Delta_\phi=\frac{d-2}2+O(m^{-1})$, but with the bilinear scalar in
the $X$ representation replaced by a Hubbard--Stratonovich field $X$ of
dimension $\Delta_X=2+O(m^{-1})$. The framework of
\cite{Henriksson:2020fqi} will then show what operator dimensions, in the
large $m$ expansion, are consistent with these assumptions.

Let us briefly review the method of large spin perturbation theory, which in the present case will be a perturbation of mean field theory, i.e. we assume that each representation $R$ in the tensor product $V\otimes V$
contains operators of spin $\ell$ and scaling dimensions
$2\Delta_\phi+2k+\ell+O(m^{-1})$. The operators with $k>0$ have OPE
coefficients that are suppressed in the $1/m$ expansion,\footnote{This
follows immediately from the expression for the mean field theory OPE coefficients
derived in \cite{Fitzpatrick:2011dm}, upon inserting
$\Delta_\phi=\frac{d-2}2+O(m^{-1})$.} and we can therefore focus on the
leading twist operators, which we denote by $R_\ell$. These are bilinear
operators of the schematic form $\phi\lsp\partial^\ell\phi$ and acquire
individual anomalous dimensions
\begin{equation}
\Delta_{R_\ell}=2\Delta_\phi+\ell+\gamma_{R_\ell},
\end{equation}
where $\gamma_{R,\ell}$ is of order $O(m^{-1})$. Symmetry under $x_1\leftrightarrow x_2$ constrains the leading twist operators such that those in the $S,X,Y,Z$ representations have even spin, and those in the $A,B$ representations have odd spin.

In large spin perturbation theory, the OPE coefficients $\lambda^2_{\phi\phi R_\ell}$ and the anomalous dimensions $\gamma_{R_\ell}$ of spinning operators in the direct channel are computed using the Lorentzian inversion formula \cite{Caron-Huot:2017vep}. The integrand of the inversion formula is proportional to the double-discontinuity $\mathrm{dDisc}[\mathcal G_R(u,v)]$, defined as the difference between the correlator and its two analytic continuations around $v=0$. In the limit $u\ll v\ll 0$, and in an expansion in $m^{-1}$, the double-discontinuity can be computed from crossed-channel operators, i.e. those appearing in the conformal block decomposition of the right-hand side of \eqref{eq:crossing}. The contribution to the $R_\ell$ from an operator $\O$ in the $\tilde R$ representation is proportional to
\begin{equation}\label{eq:dDiscO}
\mathrm{dDisc}|_\O\sim M_{R\tilde R}\lsp\lambda^2_{\phi\phi\O}\lsp \sin^2\left[\tfrac\pi 2(\tau_\O-2\Delta_\phi)\right],
\end{equation} where the argument of the squared sine is derived from the
$v\to0$ scaling of $v^{-\Delta_\phi}g_{\Delta_\O,\ell_\O}(v,u)\sim
v^{\frac12\tau_\O-\Delta_\phi}$ with $\tau_\O=\Delta_\O-\ell_\O$. The appearance of the $\sin^2$
factor means that the contribution from mean field theory operators will be
suppressed by their squared anomalous dimension.

In order to apply the framework of \cite{Henriksson:2020fqi}, we assume
that the operator $X$ has an OPE coefficient of the form
\begin{equation}\label{eq:lambdaXform}
  \lambda^2_{\phi\phi X}=\frac{a_X}{m}+O(m^{-2})\,,
\end{equation}
for some constant $a_X$ depending on $n$ and spacetime dimension $d$. The
contribution to the CFT data in the direct channel is then computed using
the inversion formula from double-discontinuities \eqref{eq:dDiscO} of
crossed-channel operators. The identity operator $\1$ generates the leading OPE
coefficients of $R_\ell$, and the operator $X$ gives a leading order
contribution to $\gamma_{R_\ell}$ in all representations $R$. In the
representations $Y,Z,A,B$ the crossed-channel operators $\1$ and $X$ provide the only contributions
at order $m^{-1}$, and, using the formulas given in \cite{Henriksson:2020fqi}, we
can write down the scaling dimensions\footnote{We present only the values
in $d=3$ dimensions, results for generic $d$ are given in
Appendix~\ref{sec:app}.}
\begin{align}\label{eq:deltaAYell}
  &\Delta_{Y_\ell}=\Delta_{A_\ell}=\ell+2\Delta_\phi-\frac{a_X}{2(\ell+1/2)(\ell-1/2)m}+O(m^{-2})\,,
\\
&\Delta_{Z_\ell}=\Delta_{B_\ell}=\ell+2\Delta_\phi+\frac{a_X}{2(\ell+1/2)(\ell-1/2)(n-1)m}+O(m^{-2})\,.
\end{align}
Recall that the spin $\ell$ is even for $Y$ and $Z$, and odd for $A$ and $B$.

Due to the combined $m$ dependence of all three factors in
\eqref{eq:dDiscO}, the anomalous dimensions in the $S$ and $X$
representations will get leading order contributions from $X$ as well as
from the $R_\ell$ in the other four representations. In the language of
\cite{Henriksson:2020fqi}, we therefore have group $\mathrm I=\{Y,Z,A,B\}$,
and group $\mathrm{II}=\{S,X\}$. Evaluating the formulas of that paper
gives
\begin{align}
&\Delta_{S_\ell}=\ell+2\Delta_\phi-\frac{a_X}{2(\ell+1/2)(\ell-1/2)m}-\frac{
\pi^2\,\ell\,a_X^2n}{4(\ell+1/2)(\ell-1/2)(n-1)m}+O(m^{-2})\,,
\\\label{eq:deltaXell}
&\Delta_{X_\ell}=\ell+2\Delta_\phi-\frac{a_X}{2(\ell+1/2)(\ell-1/2)m}-\frac{
\pi^2\,\ell\,a_X^2n(n-2)}{4(\ell+1/2)(\ell-1/2)(n-1)^2m}+O(m^{-2})\,.
\end{align}
The expressions \eqref{eq:deltaAYell}--\eqref{eq:deltaXell} depend on two
unknowns: the constant $a_X$ introduced in \eqref{eq:lambdaXform}, and the
leading anomalous dimension $\gamma_\phi^{(1)}$ defined by
$\Delta_\phi=\frac{d-2}2+\gamma^{(1)}_\phi/m+O(m^{-2})$. However,
conservation of the global symmetry current and the stress-energy tensor
gives the two equations $\Delta_{A_1}=d-1$, $\Delta_{S_2}=d$. The latter
equation is quadratic in $a_X$, and we get two solutions:
$a_X=\gamma^{(1)}_\phi=0$ and
\begin{equation}
  a_X=\frac{4 (n-1)}{\pi ^2 n}\,,\qquad
  \gamma_\phi^{(1)}=\frac{4 (n-1)}{3 \pi^2 n}\,.
\end{equation}
Choosing the non-trivial solution, we have fixed the leading order
anomalous dimensions of all leading twist spinning operators in the theory.

We have also computed the corrections to the OPE coefficients of these
operators, defined with respect to the leading order result
given by $M_{RS}$ times the OPE coefficients of mean field theory. In general, these
results are not particularly illuminating, but specifying to the conserved
operators we extract the corrections to the central charge and the current
central charge,\footnote{The observant reader may note that these results
agree with those of one free and $n-1$ interacting $O(m)$ models. This
agreement is broken at higher orders in $m^{-1}$, which can be seen from
the $\varepsilon$ expansion, where
$C_T/C_{T,\mathrm{free}}=1-5\gamma^{(2)}_\phi\varepsilon^2/3+O(\varepsilon^3)$
\cite{Henriksson:2020fqi}.}
\begin{align}
\frac{C_T}{C_{T,\mathrm{free}}}&=1-\frac{40(n-1)}{9n}\frac{1}{\pi^2m}+O(m^{-2})\,,
\\
\frac{C_J}{C_{J,\mathrm{free}}}&=1-\frac{64(n-1)}{9n}\frac{1}{\pi^2m}+O(m^{-2})\,.
\end{align}

The most important class of observables is the dimensions of the leading
scalar operators in each representation. Unfortunately, spin zero is beyond
the guaranteed region of convergence of the Lorentzian inversion formula
and it is not \emph{a priori} clear how to extract these values. However,
in the large $N$ expansion of theories with $O(N)$ \cite{Alday:2019clp} and
$O(m)\times O(N)$ \cite{Henriksson:2020fqi} global symmetry, it has been
observed that evaluating $\Delta_{R_\ell}$ for $\ell=0$ correctly
reproduces the dimensions of the scalar operators as computed by
independent methods.  If we assume that this is the case also for our \MN
symmetric theory, we can find the scalar operator dimensions by demanding
that $\Delta_X=d-\Delta_{X_{\ell=0}}$ and that $\Delta_R=\Delta_{R_{\ell=0}}$
for $R=S,Y,Z$. Further support for this assumption is that the expressions
derived from this assumptions, evaluated for $d=4-\varepsilon$, agree with
the expressions \eqref{eq:deltaS}--\eqref{eq:deltaZ} in the overlap of the
orders: $\varepsilon^3/m$.  In three dimensions we find the following
dimensions of the scalar operators:
\begin{align}\label{eq:mdeltaphi}
  \Delta_\phi&=\frac12+\frac{4(n-1)}{3n}\frac1{\pi^2m}+O(m^{-2})\,,
\\\label{eq:mdeltaS}
\Delta_S&=1+\frac{32(n-1)}{3n}\frac1{\pi^2m}+O(m^{-2})\,,
\\\label{eq:mdeltaX}
\Delta_X&=2-\frac{32(n-1)}{3n}\frac1{\pi^2m}+O(m^{-2})\,,
\\\label{eq:mdeltaY}
\Delta_Y&=1+\frac{32(n-1)}{3n}\frac1{\pi^2m}+O(m^{-2})\,,
\\\label{eq:mdeltaZ}
\Delta_Z&=1+\frac{8(n-4)}{3n}\frac1{\pi^2m}+O(m^{-2})\,,
\end{align}
and results for generic spacetime dimension $d$ are presented in
Appendix~\ref{sec:app}.

\subsec{Comparison with numerical results}\label{sec:comparenum}

\begin{figure}[ht]
  \centering
  \includegraphics{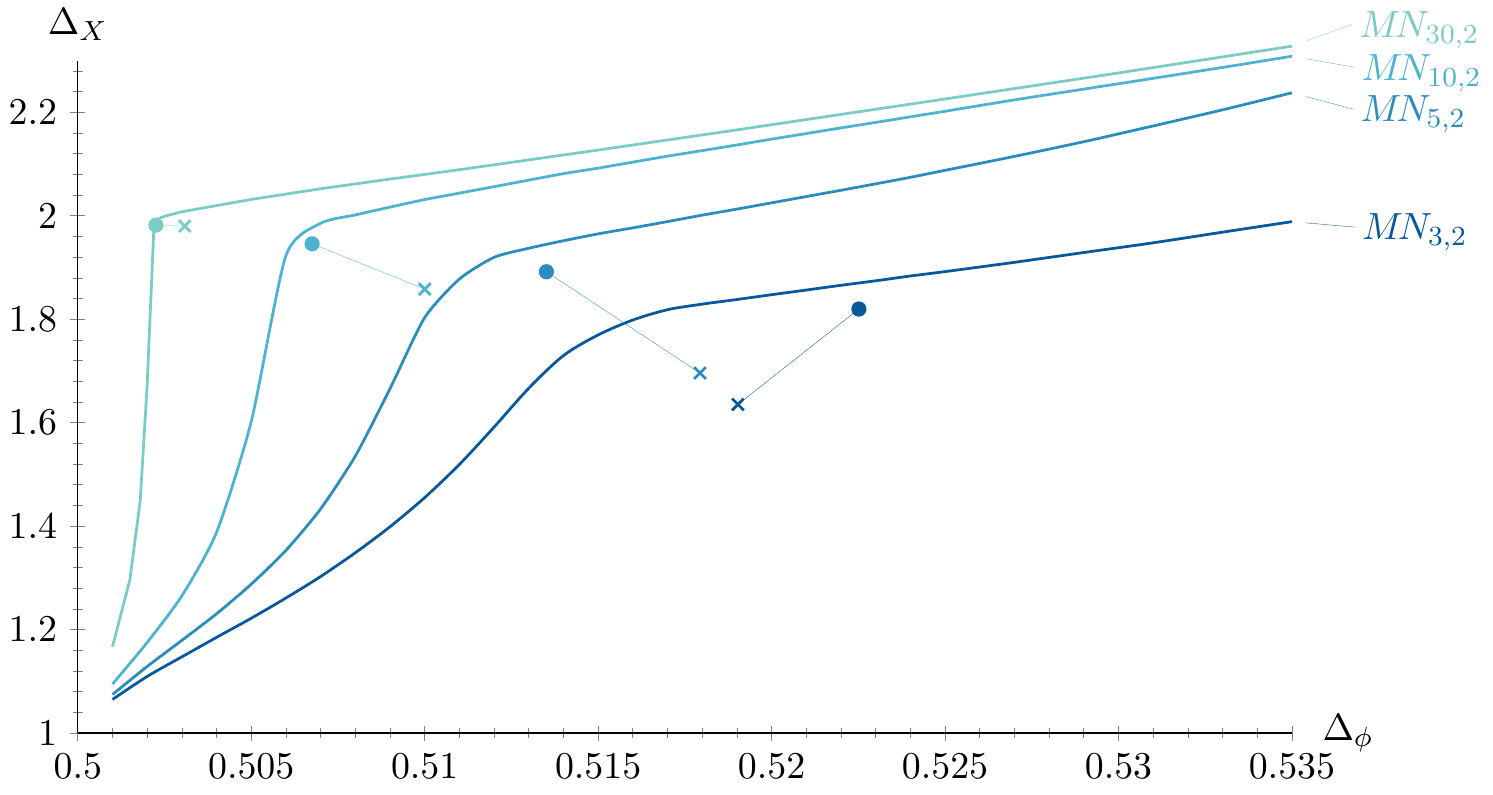}
  \caption{Bounds and corresponding locations of fixed points given as dots
  for large $m$ and crosses for $\veps$ expansion results. The lines
  connecting dots and crosses are drawn to help illustrate results pertaining to
  the same theory.}
  \label{fig:kink-1-multi}
\end{figure}

The results \eqref{eq:mdeltaphi}--\eqref{eq:mdeltaZ} can now be compared
with the numerical bootstrap results for the first kink. In
Fig.~\ref{fig:kink-1-multi} we display the bounds in the
$(\Delta_\phi,\Delta_X)$ plane from Fig.~2 of \cite{Stergiou:2019dcv},
together with our new large $m$ results \eqref{eq:mdeltaphi},
\eqref{eq:mdeltaX}, as well as the $\varepsilon^3$ results
\eqref{eq:deltaphi}, \eqref{eq:deltaX}\footnote{In producing this graph, as
well as the corresponding values in Table~\ref{table:comparison-kink1}, we have used the
truncated results of the $\varepsilon$ expansion at order $\varepsilon^3$. We comment on this in
section~\ref{sec:epsilonexpansionQ}.} from the literature.
We see that the agreement is good between all three methods for $m\gg1$, and that the large $m$ expansion better captures the finite $m$ behavior than does the $\varepsilon$ expansion.

Our new results in the large $m$ expansion can also be used to derive predictions for the critical exponents using the relations
\begin{align}\label{eq:exponents1}
  \eta &= 2-d+2\Delta_\phi\,,&
  \nu^{-1}&=d-\Delta_S\,,&\alpha&=2-\nu d\,,
\\\label{eq:exponents2}
\beta &=\nu\,\Delta_\phi\,,&
\gamma &=\nu\,(d-2\Delta_\phi)\,,&\phi_\kappa&=\nu(d-\Delta_Z)\,.
\end{align}
For the physically relevant case $\MNmath_{2,2}$, our new predictions for $\beta$ and $\nu$ from the large $m$ expansion are closer to numerical bootstrap results, as well as experiments and Monte Carlo results, than predictions from the $\varepsilon$ expansion, as can be seen in Table~\ref{table:comparison-kink1}. Also for the chiral cross-over exponent $\phi_\kappa$, the value $1.2343$ derived from our large $m$ results\footnote{We first estimated $\Delta_S$ and $\Delta_Z$ by evaluating the truncated expansions \eqref{eq:mdeltaS} and \eqref{eq:mdeltaZ} for $m=n=2$, before using \eqref{eq:exponents2} to find $\phi_\kappa$.} compares favorably with the Monte Carlo value  $1.22(6)$ \cite{Kawamura1992}.

\begin{table}[ht]
\centering
\caption{Comparison of data for \MN symmetric theories across various methods. The truncated series denote truncation to orders $\varepsilon^3$
  and $m^{-1}$ respectively for the scaling dimensions. These numerical
  values are then used in \eqref{eq:exponents1} and \eqref{eq:exponents2} to give estimates for $\beta$ and $\nu$.
}\label{table:comparison-kink1}
\begin{tabular}{|l|ccccc|}
\hline
{$\boldsymbol{\MNmath_{2,2}}$}&$\Delta_\phi$&$\Delta_S$&$\Delta_X$& $\beta$ &$\nu$
\\\hline

XY STA &&& & $0.24(2)$& $0.55(5)$
\\
Tb &&& & $0.23(4)$& $0.53(4)$
\\
Monte Carlo \cite{Kawamura1992} &&&& $0.253(10)$ & $0.54(2)$
\\
Monte Carlo \cite{Calabrese:2004nt} &&&& $0.317(35)$ & $0.63(7)$
\\\hdashline
Numerical bootstrap \cite{Stergiou:2019dcv} 
& 0.518(1) &1.233(20)&
1.676(10)& $0.293(3)$ & $0.566(6)$
\\
$\varepsilon^4$ resummation \cite{Mudrov:2001yr,Mudrov2} &&&&  $ 0.370(5)$ &  $0.715(10)$
\\
$\varepsilon$ expansion (trunc.)& $0.5191$ & $1.5052$ & $1.3808$ & $0.3473$ & $0.6690$
\\
Large $m$ expansion (trunc.) &$0.5338$&$1.2702$&$1.7298	$&$0.3086 $& $0.5781$
\\
\hline
{$\boldsymbol{\MNmath_{2,3}}$}&$\Delta_\phi$&$\Delta_S$&$\Delta_X$& $\beta$ &$\nu$
\\\hline
Numerical bootstrap \cite{Stergiou:2019dcv}
 &0.518(1)&1.279(20)&1.590(10)&$0.301(3)$& $0.581(6)$
\\
$\varepsilon^4$ resummation \cite{Mudrov:2001yr,Mudrov2} &&&&  $ 0.363(6)$ &  $0.702(10)$
\\
$\varepsilon$ expansion (trunc.) & $0.5186$ & $1.4642$ & $1.3976$ & $ 0.3377$ & $0.6511$
\\
Large $m$ expansion (trunc.) &$0.5450$&$1.3603$&$1.6398$&$0.3324$& $0.6099$
\\
\hline
{$\boldsymbol{\MNmath_{20,2}}$}&$\Delta_\phi$&$\Delta_S$&$\Delta_X$& $\beta$ &$\nu$
\\\hline
Numerical bootstrap &0.5032(1)&1.025(20)&1.965(10)&0.2548(26)&0.506(5)
\\
$\varepsilon$ expansion (trunc.) & $0.5047$&$1.0215$&$1.9606$&$0.2551$&  $0.5054$
\\
Large $m$ expansion (trunc.)  & $0.5034$ & $1.0270$ & $1.9730$ &  $0.2551$ &  $0.5068$
\\
\hline
\end{tabular}
\end{table}

\subsec{Connection to the \texorpdfstring{$\varepsilon$}{epsilon} expansion}\label{sec:epsilonexpansionQ}

As we mentioned in the introduction, results derived in the $\varepsilon$
expansion have not been successful in matching the experimental values
observed in the cases $\MNmath_{2,2}$ and $\MNmath_{2,3}$. This is in
contrast to critical phenomena described by CFTs with several other
symmetry groups, where the results in the $\varepsilon$ expansion give
surprisingly good agreement with experimental data as well as
non-perturbative results from Monte Carlo simulations and numerical conformal bootstrap.

The lack of agreement between bootstrap and $\varepsilon$ expansion results
in the $\MNmath_{2,2}$ case may be taken as a sign that the fixed point
found of the $\varepsilon$ expansion, as discussed in
section~\ref{sec:epsresults}, is not connected to the CFT describing the
critical phenomena in three dimensions. Our results strongly indicate the
contrary, and that the connection is manifest through the large $m$
expansion derived above. Specifically, near four dimensions our new
analytic results agree with the $\varepsilon$ expansion, and in three
dimensions, the large $m$ expansion is connected to the finite $m$ CFTs
through the family of kinks displayed in Fig.~\ref{fig:kink-1-multi}.

We note that for the larger values of $m$ there is good agreement between
all three methods: large $m$ expansion, $\varepsilon$ expansion, and
numerical conformal bootstrap. For the lower values of $m$, our new large
$m$ expansion evaluates at a point closer to the corresponding kink than
does the $\varepsilon$ expansion. For the latter, we have simply used
direct truncation of the order $\varepsilon^3$ results; alternatively one
could use Pad\'e approximants or various resummation techniques.\footnote{
We find that the Pad\'e approximants constructed from the order
$\varepsilon^3$ results contain spurious poles in the region
$\varepsilon\leqslant 1$, and since the $\varepsilon$ expansion is not the
focus of this paper we have not attempted any resummation methods. Note
that the resummed $\varepsilon$ expansion of \cite{Mudrov2}, included in
Table~\ref{table:comparison-kink1}, does not give any improvement compared
to a direct truncation.} In Table~\ref{table:comparison-kink1} we extended
this comparison to more observables, and again we get an improved agreement
with the numerical bootstrap compared to the $\varepsilon$ expansion. Note,
for instance, that for small $m$ the $\varepsilon $ expansion predicts that
$\Delta_X<\Delta_S$, which is inconsistent with the bootstrap results.

In Table~\ref{table:comparison-kink1}, for the $\MNmath_{2,2}$ case, we have also included some results from experiments and Monte Carlo simulations, which in the literature are assigned to the chiral universality class. As mentioned in the introduction, the $\MNmath_{2,2}$ fixed point
obtained in the $\varepsilon$ expansion has $g<0$ in \LagMN, which means
that it should not be applicable to these cases. However, $g<0$ in the
$\varepsilon$ expansion does not guarantee that $g<0$ as $\varepsilon$ becomes finite, and therefore it cannot be ruled out that at $\varepsilon=1$ the fixed point of kink 1 in fact has $g>0$ and thus lies in the chiral region. Unfortunately, we are
currently unable to probe the sign of $g$ using our
bootstrap methods.

While the values for the critical exponent $\nu$ show good agreement across
experiments, bootstrap, Monte Carlo and large $m$ expansion, the situation
for the exponent $\beta$ is more concerning. In fact, as already pointed
out in the literature~\cite{Pelissetto:2000ek,Delamotte:2003Dw}, some of the
experimental and Monte Carlo values do not satisfy the constraint
$2\beta-\nu\geqslant 0$ as implied by the unitarity bound \eqref{eq:UB} for
$\Delta_\phi$.  This inconsistency could be explained by unknown systematic
errors of these methods, or that they are not measuring the exponents at
criticality. We do not wish to comment further on this, more than to point
out that our results give exponents consistent with unitarity.

\newsec{The non-perturbative fixed point at large \texorpdfstring{$\boldsymbol{m}$}{m}}\label{sec:kink2}

In this section we study, using the numerical bootstrap, the large $m$
limit of the second kink for $\MNmath_{m,n}$ symmetric theories.  While two
kinks are clearly visible for $\MNmath_{2,3}$ in
\cite[Fig.~4]{Stergiou:2019dcv}, we find that the second kink only persists
at large $m$ for the case $n=2$, and we focus our attention to the cases
$\MNmath_{m,2}$ for various $m$.

\subsec{Numerical bootstrap study}\label{sec:numericalimplementation}

\begin{figure}[H]
  \centering
  \includegraphics[scale=0.9]{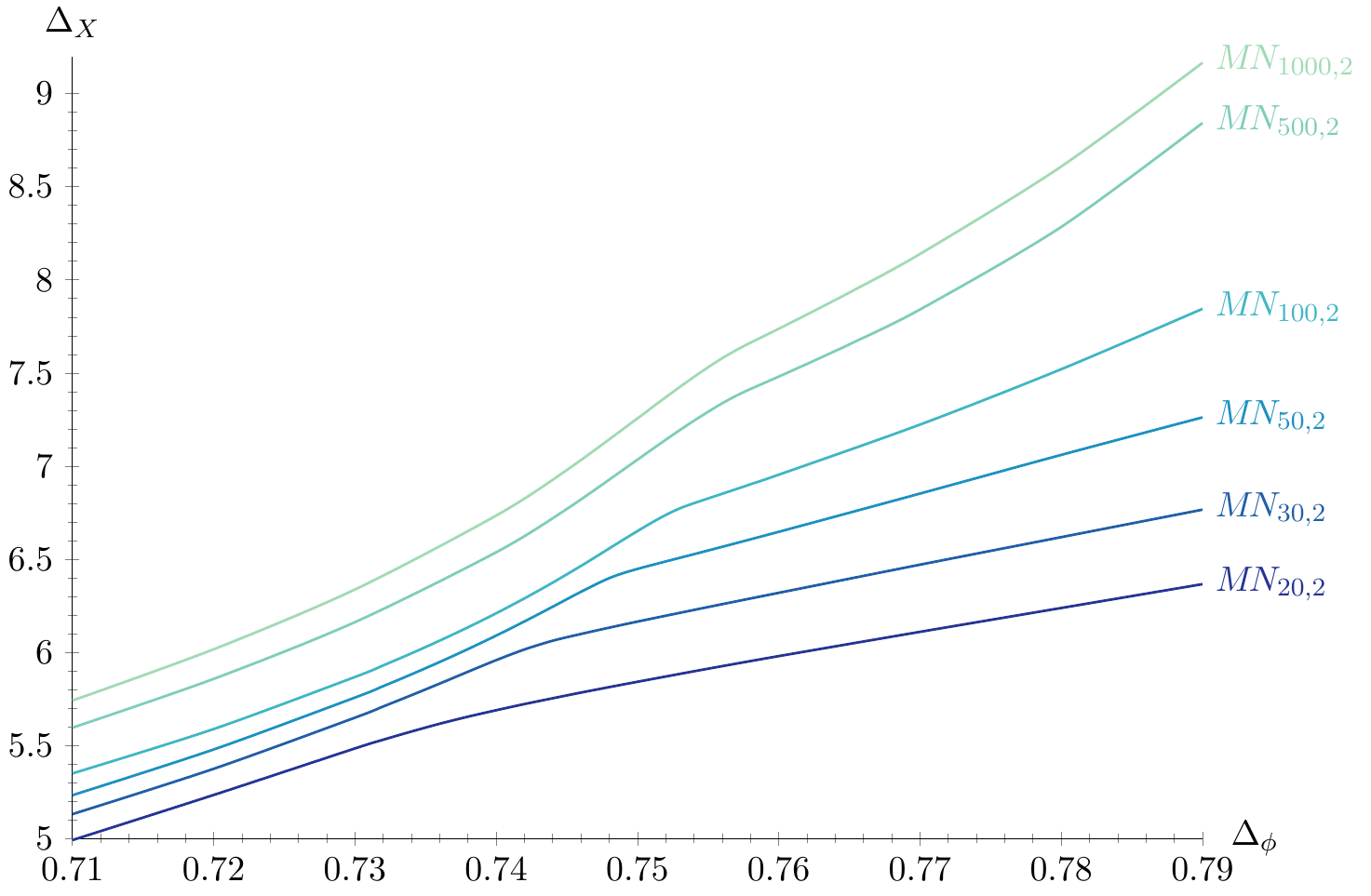}
  \caption{Bounds for 3D CFTs with $\MNmath_{m,2}$ global symmetry for various values of $m$. The allowed region is below the curves in the corresponding
  theories. These bounds are obtained with the use of \texttt{PyCFTBoot}
  \cite{Behan:2016dtz} with parameters $\texttt{n\_max=9}$,
  $\texttt{m\_max=6}$, $\texttt{k\_max=36}$ and $\texttt{l\_max=26}$.}
  \label{fig:kink-2-multi}
\end{figure}

Our first set of results consists of bootstrap bounds for $\MNmath_{m,2}$
theories for large values of $m$; see Fig.~\ref{fig:kink-2-multi}. These
bounds show that the kink persists at large $m$ and that its position
stabilizes close to $\Delta_\phi=0.75$. However, the kink still moves
significantly in the $\Delta_X$ direction.

\begin{figure}[H]
  \centering
  \includegraphics[scale=0.9]{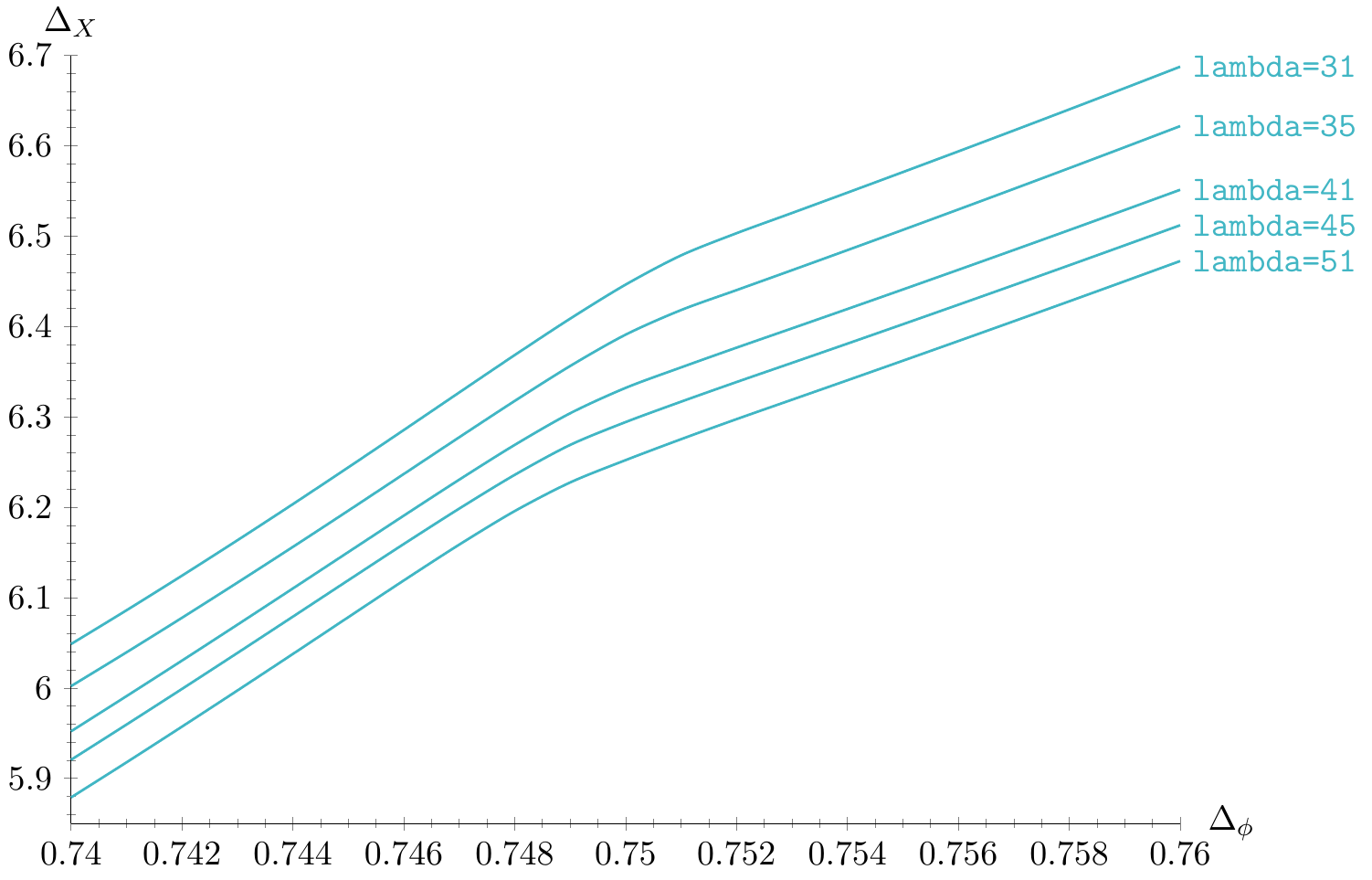}
  \caption{Bounds for 3D CFTs with $\MNmath_{100,2}$ global symmetry with
  increasing numerical strength (top to bottom). For these bounds we used
  \texttt{qboot}~\cite{Go:2020ahx}. These bounds are all stronger than the
  corresponding bounds in Fig.~\ref{fig:kink-2-multi}. The latter are
  obtained with \texttt{qboot} with the choice $\Lambda\approx 25$.}
  \label{fig:position-100-2}
\end{figure}

The position for $\Delta_X$ of the kink is not stable upon increasing the
number of derivatives; see Fig.~\ref{fig:position-100-2}. It appears,
however, that $\Delta_\phi$ is fairly stable near the value $0.75$. We
conjecture that
\begin{equation}\label{eq:deltaphiconj}
  \Delta_\phi|_{m\to\infty,n=2}=0.75(1)\,.
\end{equation}
For our strongest numerics we used \texttt{qboot}~\cite{Go:2020ahx} with
parameters $\texttt{prec=1300}$, $\texttt{n\_Max=560}$,
$\texttt{lambda=51}$, $\texttt{numax=30}$ and the set of spins \{0, \ldots,
80, 85, 86, 89, 90, 93, 94, 97, 98, 101, 102, 105, 106, 109, 110, 111, 112,
115, 116, 119, 120\}.

We also obtained bounds for $m=100$ with $n=3,4$. As we see in
Fig.~\ref{fig:Delta_X_bound_MN_100,2-100,4}, the kink is clearly present
only for $n=2$. This suggests that there exists a critical line $n_c(m)$
below which we have two distinct CFTs with $\MNmath_{m,n}$ global
symmetry.

\begin{figure}[ht]
  \centering
  \includegraphics[scale=0.9]{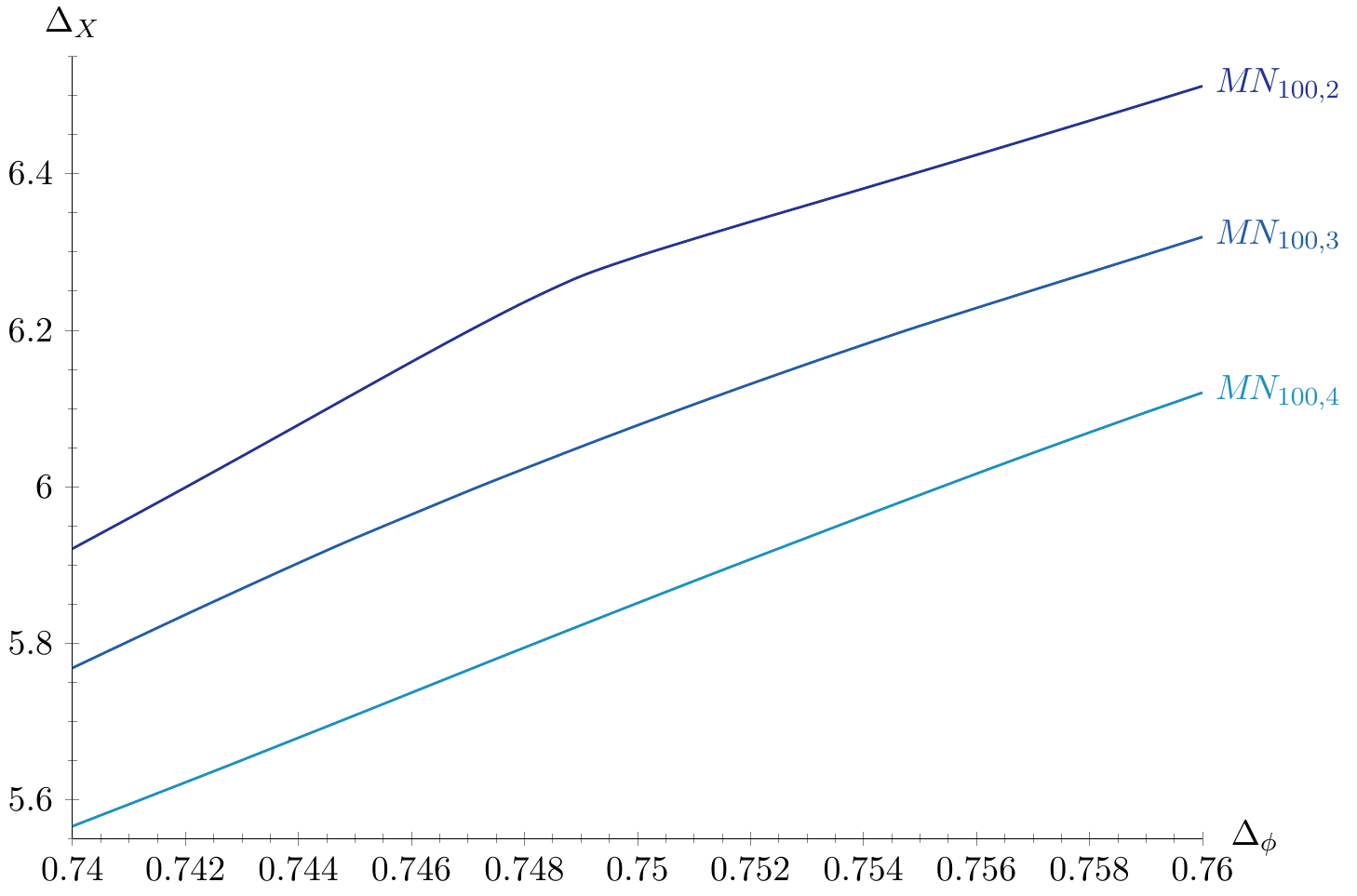}
  \caption{Bounds for 3D CFTs with $\MNmath_{100,n}$ global symmetry for various values of $n$. The allowed region is below the curves in the corresponding
  theories. For these bounds we used \texttt{qboot}~\cite{Go:2020ahx} with
  parameters $\texttt{prec=1200}$, $\texttt{n\_Max=520}$,
  $\texttt{lambda=45}$, $\texttt{numax=26}$ and the set of spins \{0,
  \ldots, 60, 63, 64, 66, 67, 73, 74, 77, 78, 81, 82, 85, 86, 89, 90, 93,
  94, 97, 98\}.}
  \label{fig:Delta_X_bound_MN_100,2-100,4}
\end{figure}

\begin{figure}
  \centering
      \begin{subfigure}[b]{0.27\textwidth}
        \vspace{7.5pt}\includegraphics[width=\textwidth]{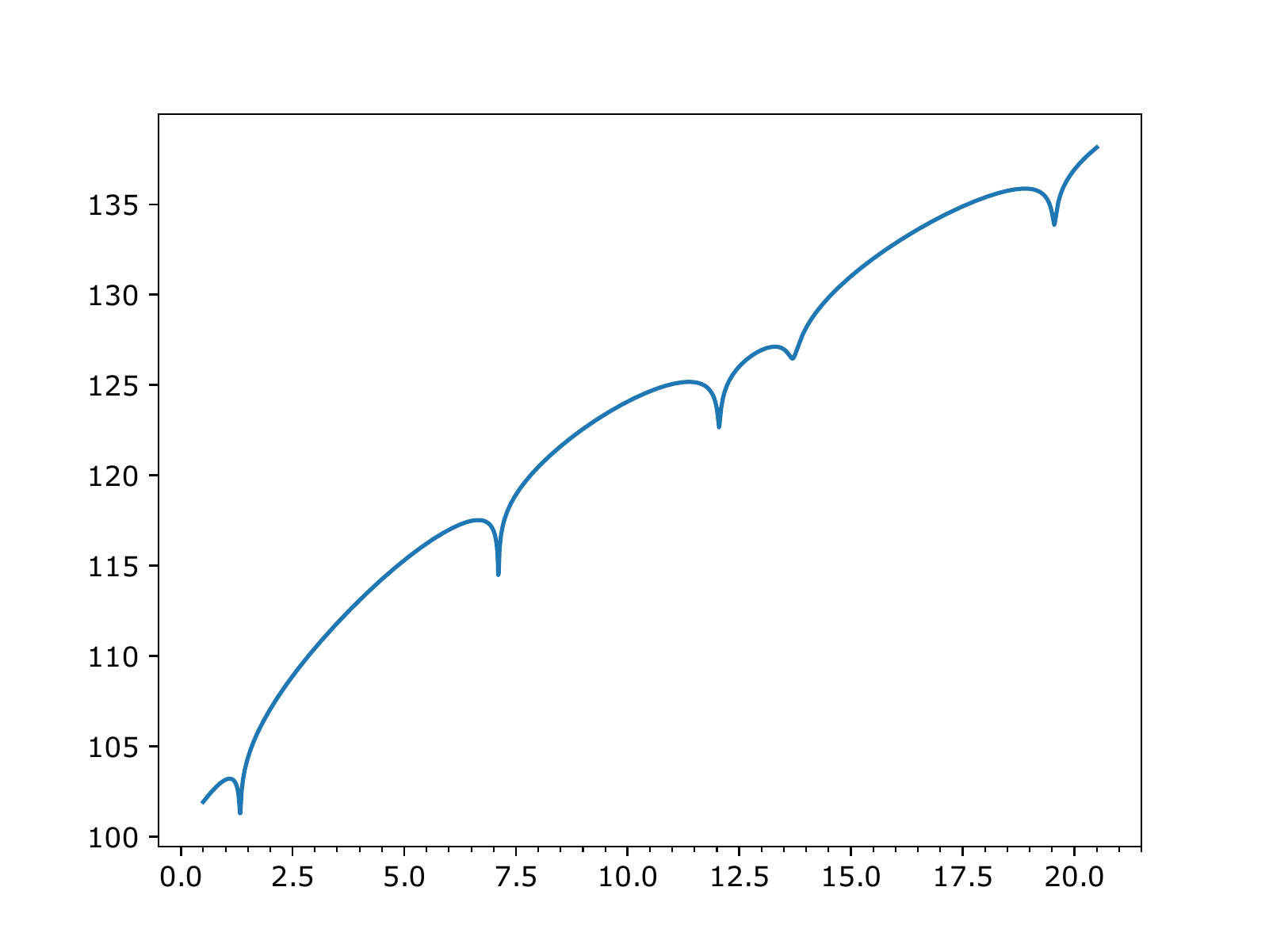}\vspace{-7.5pt}
        \caption{$S$, $\ell=0$}\label{subfig:S0}
    \end{subfigure}
    ~
          \begin{subfigure}[b]{0.27\textwidth}
        \vspace{7.5pt}\includegraphics[width=\textwidth]{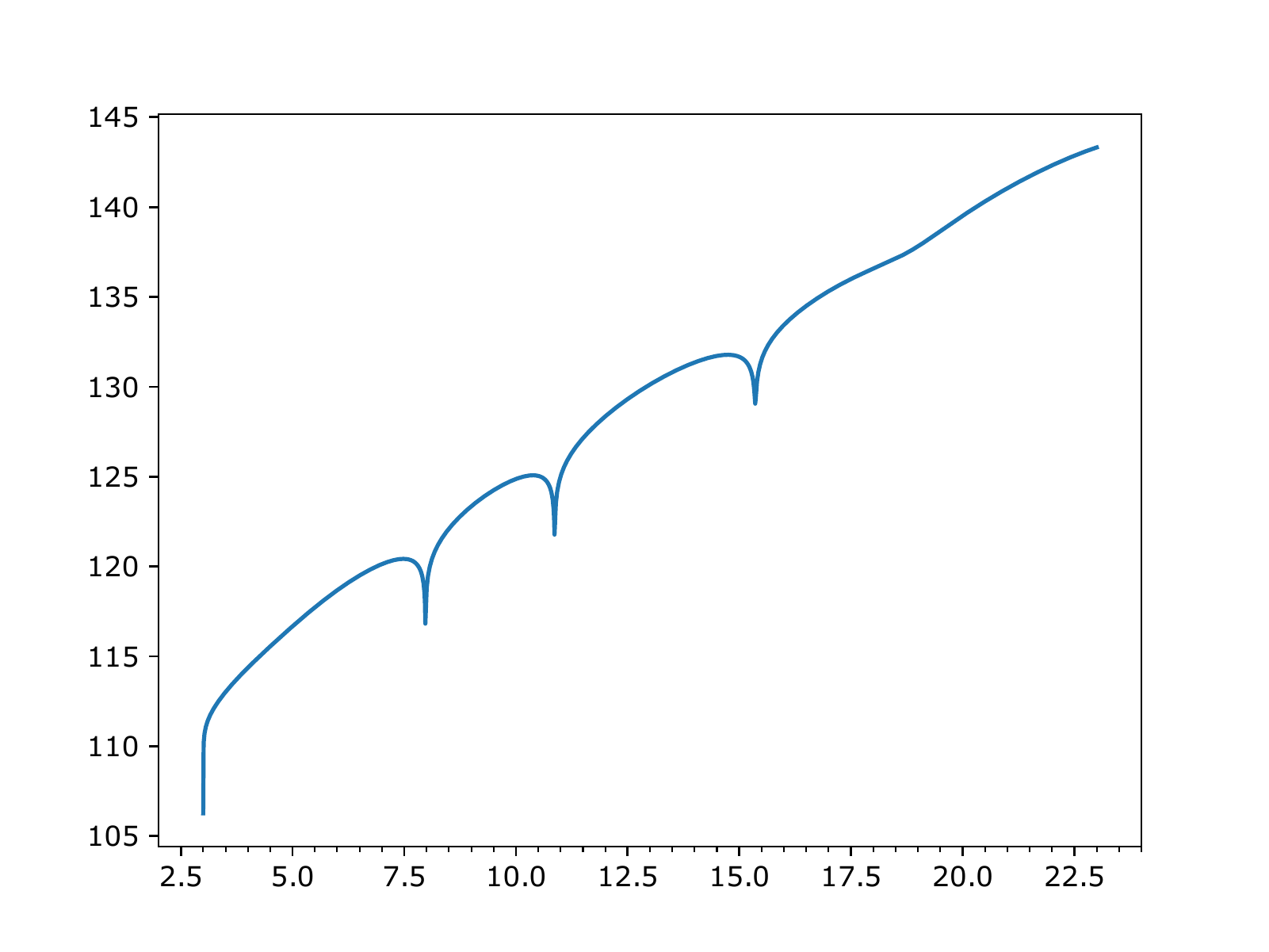}\vspace{-7.5pt}
        \caption{$S$, $\ell=2$}
    \end{subfigure}
    ~
          \begin{subfigure}[b]{0.27\textwidth}
        \vspace{7.5pt}\includegraphics[width=\textwidth]{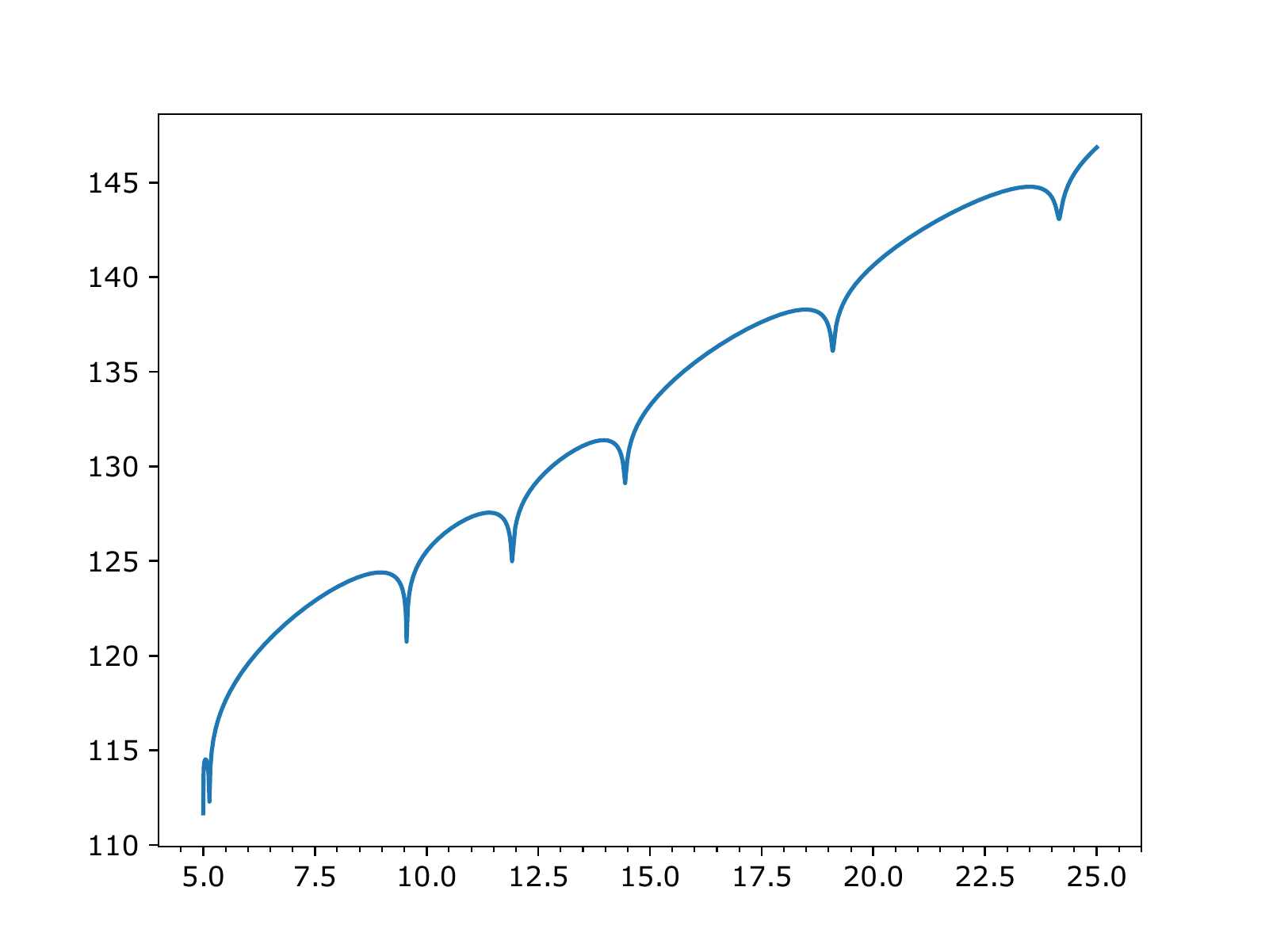}\vspace{-7.5pt}
        \caption{$S$, $\ell=4$}
    \end{subfigure}
    ~

          \begin{subfigure}[b]{0.27\textwidth}
        \vspace{7.5pt}\includegraphics[width=\textwidth]{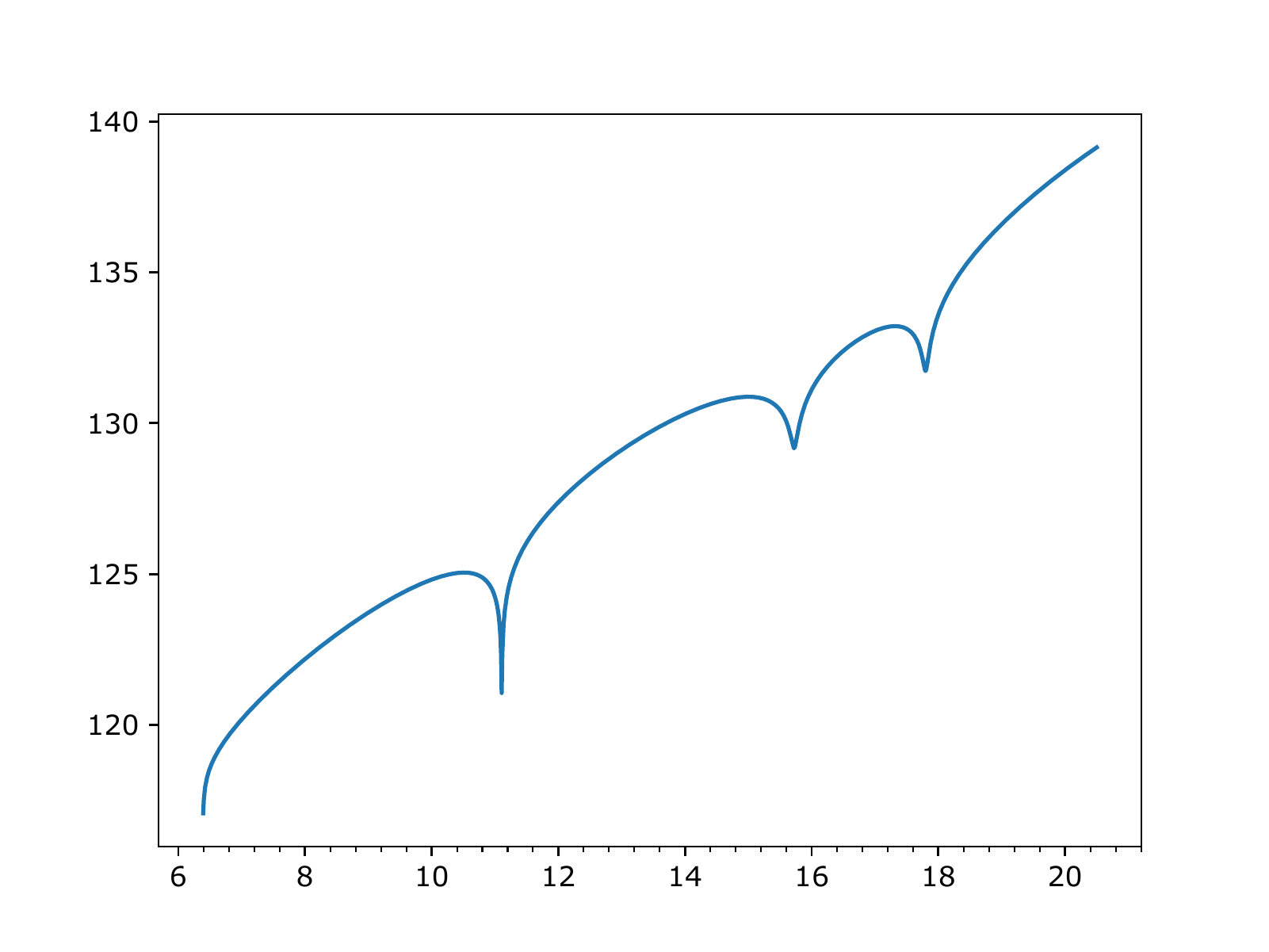}\vspace{-7.5pt}
        \caption{$X$, $\ell=0$}
    \end{subfigure}
    ~
          \begin{subfigure}[b]{0.27\textwidth}
        \vspace{7.5pt}\includegraphics[width=\textwidth]{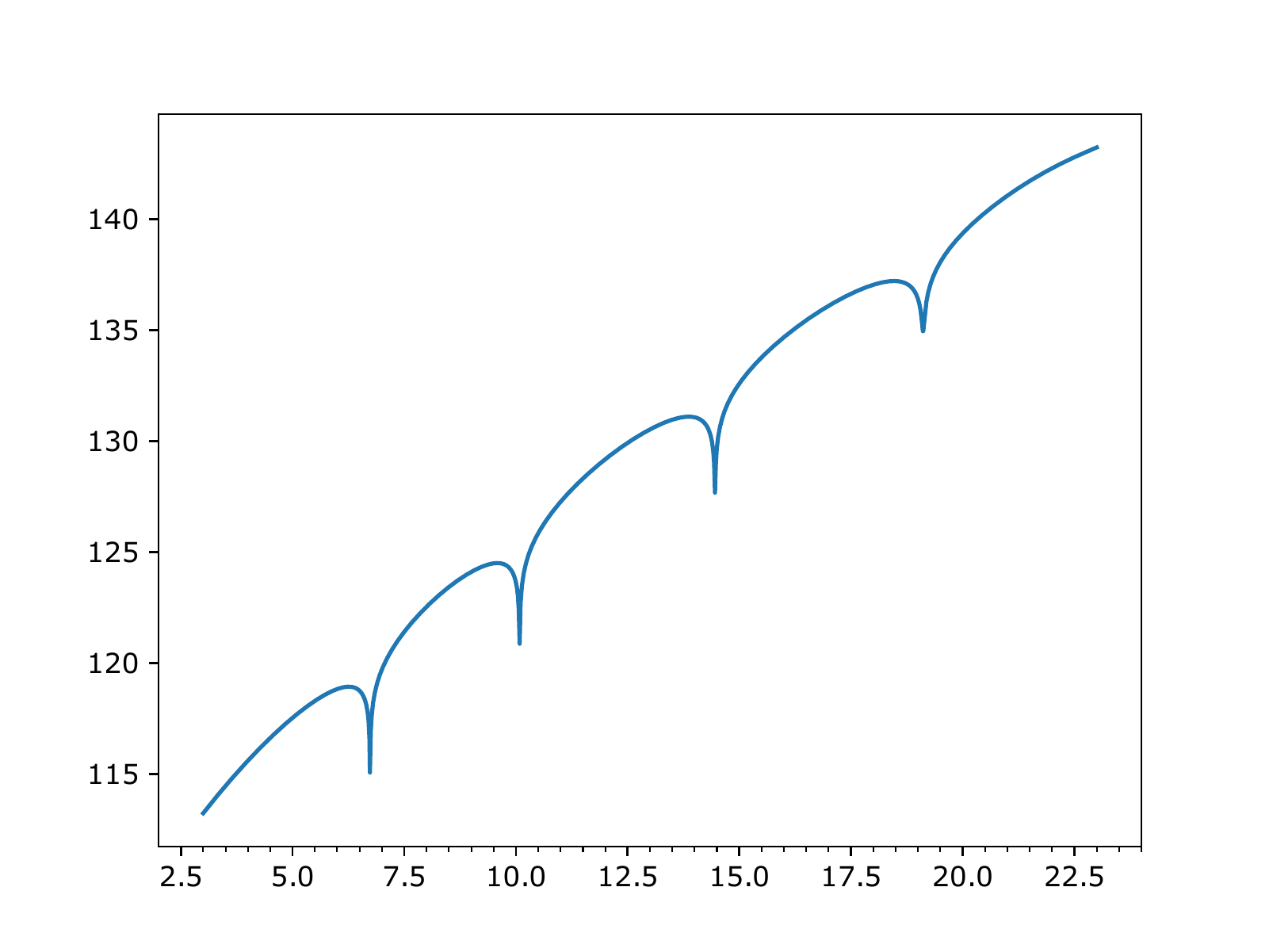}\vspace{-7.5pt}
        \caption{$X$, $\ell=2$}
    \end{subfigure}
    ~
          \begin{subfigure}[b]{0.27\textwidth}
        \vspace{7.5pt}\includegraphics[width=\textwidth]{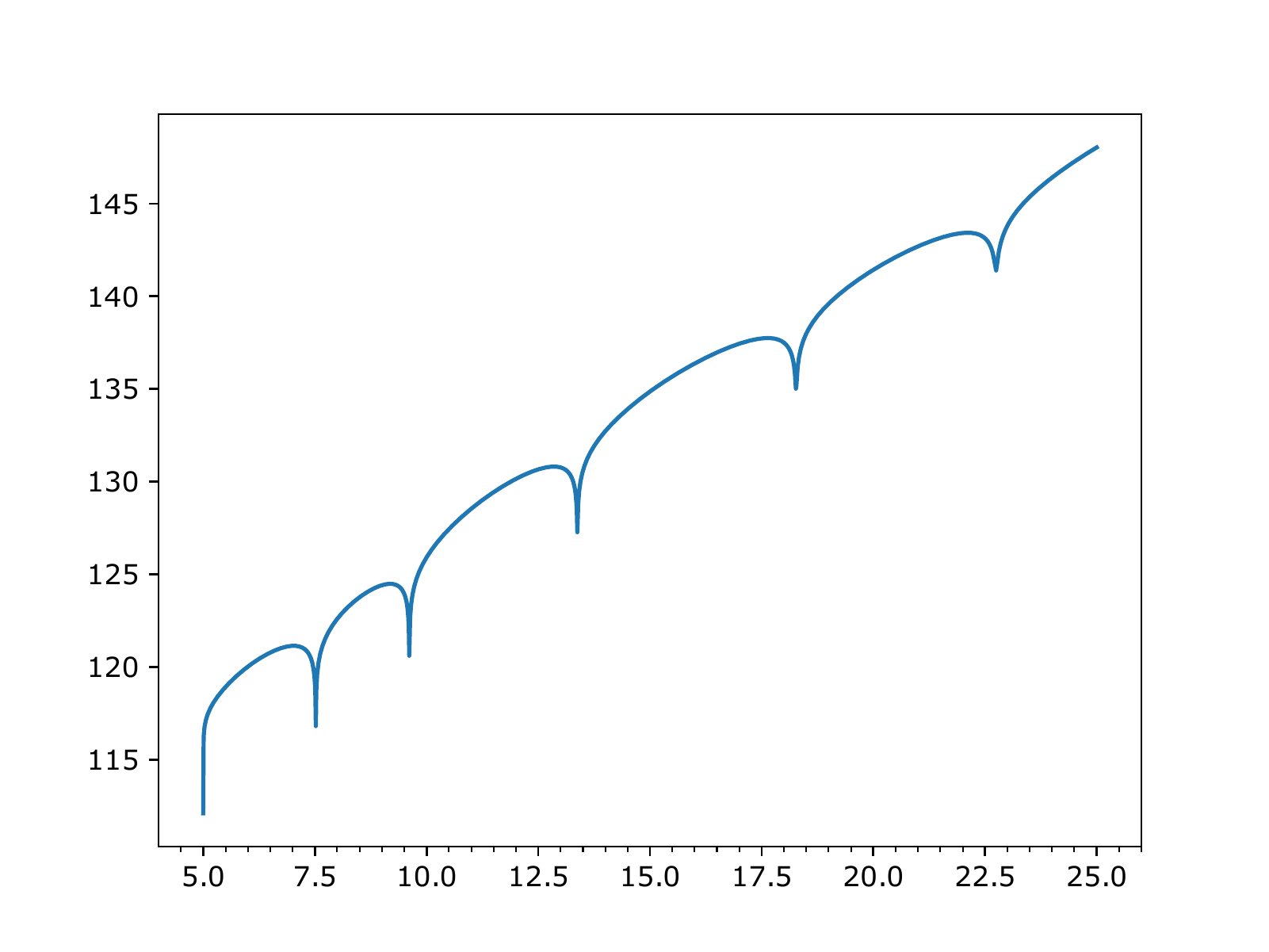}\vspace{-7.5pt}
        \caption{$X$, $\ell=4$}\label{subfig:X4}
    \end{subfigure}
    ~

          \begin{subfigure}[b]{0.27\textwidth}
       \vspace{7.5pt} \includegraphics[width=\textwidth]{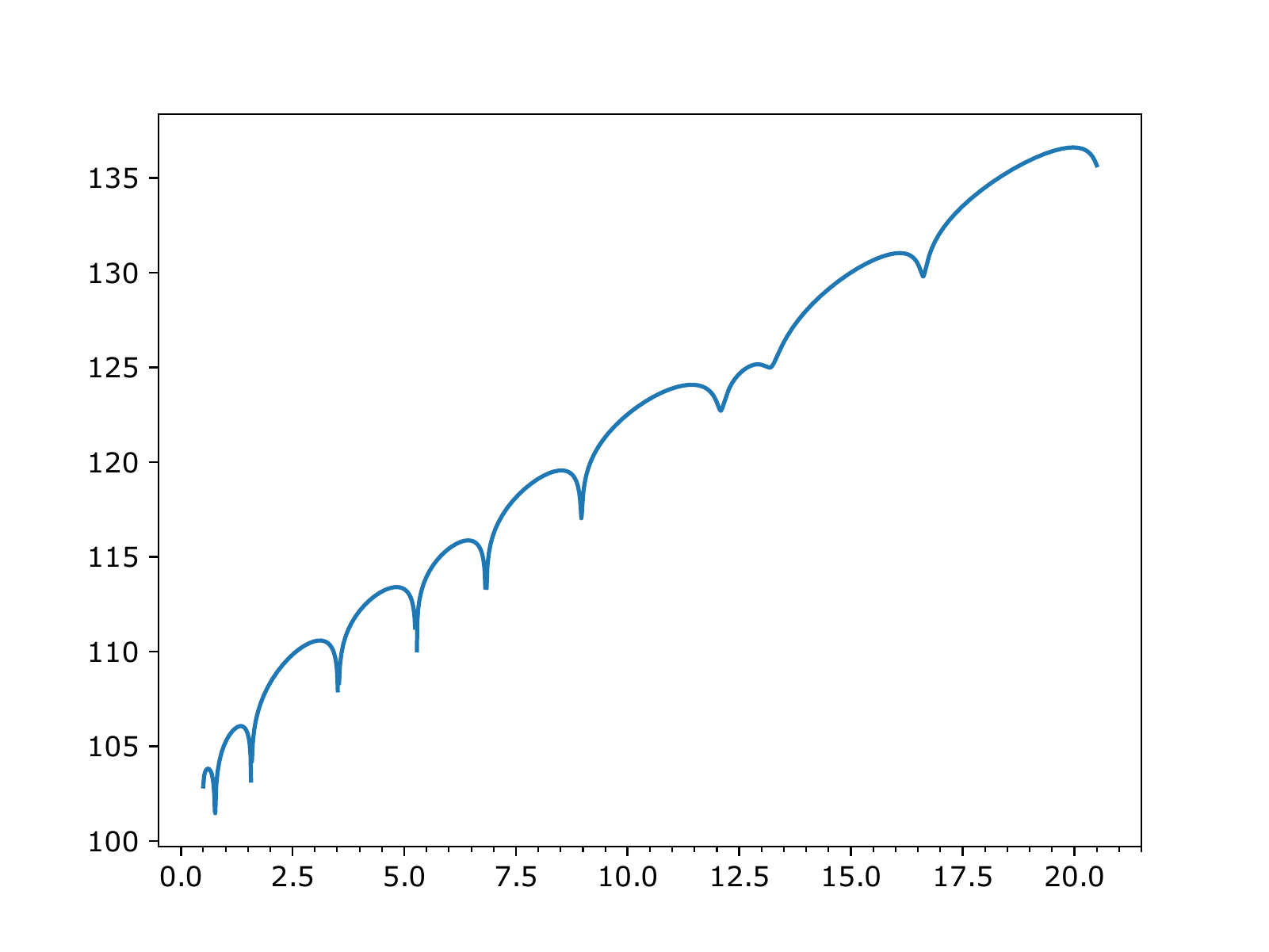}\vspace{-7.5pt}
        \caption{$Y$, $\ell=0$}
    \end{subfigure}
    ~
          \begin{subfigure}[b]{0.27\textwidth}
       \vspace{7.5pt} \includegraphics[width=\textwidth]{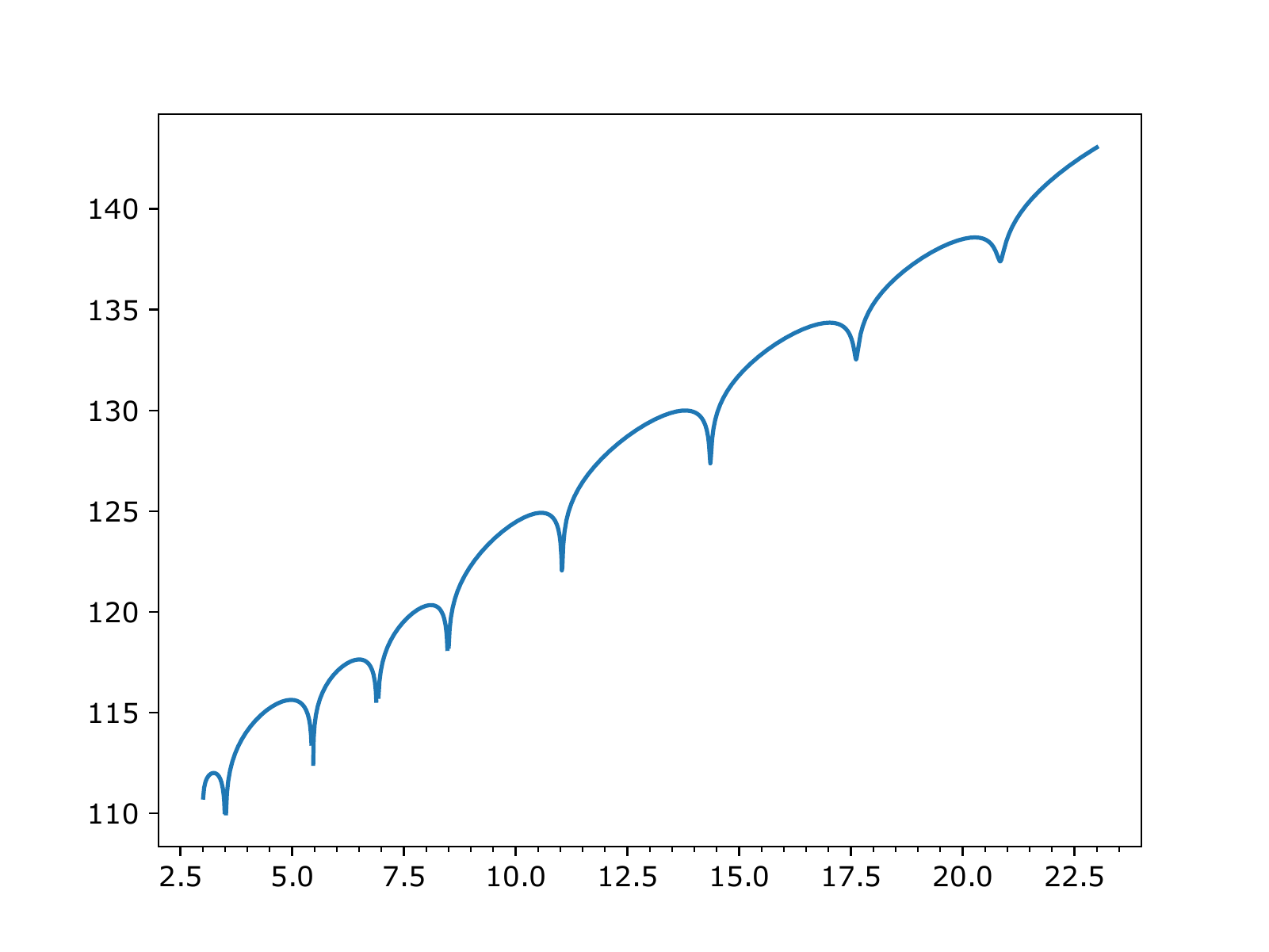}\vspace{-7.5pt}
        \caption{$Y$, $\ell=2$}
    \end{subfigure}
    ~
          \begin{subfigure}[b]{0.27\textwidth}
       \vspace{7.5pt} \includegraphics[width=\textwidth]{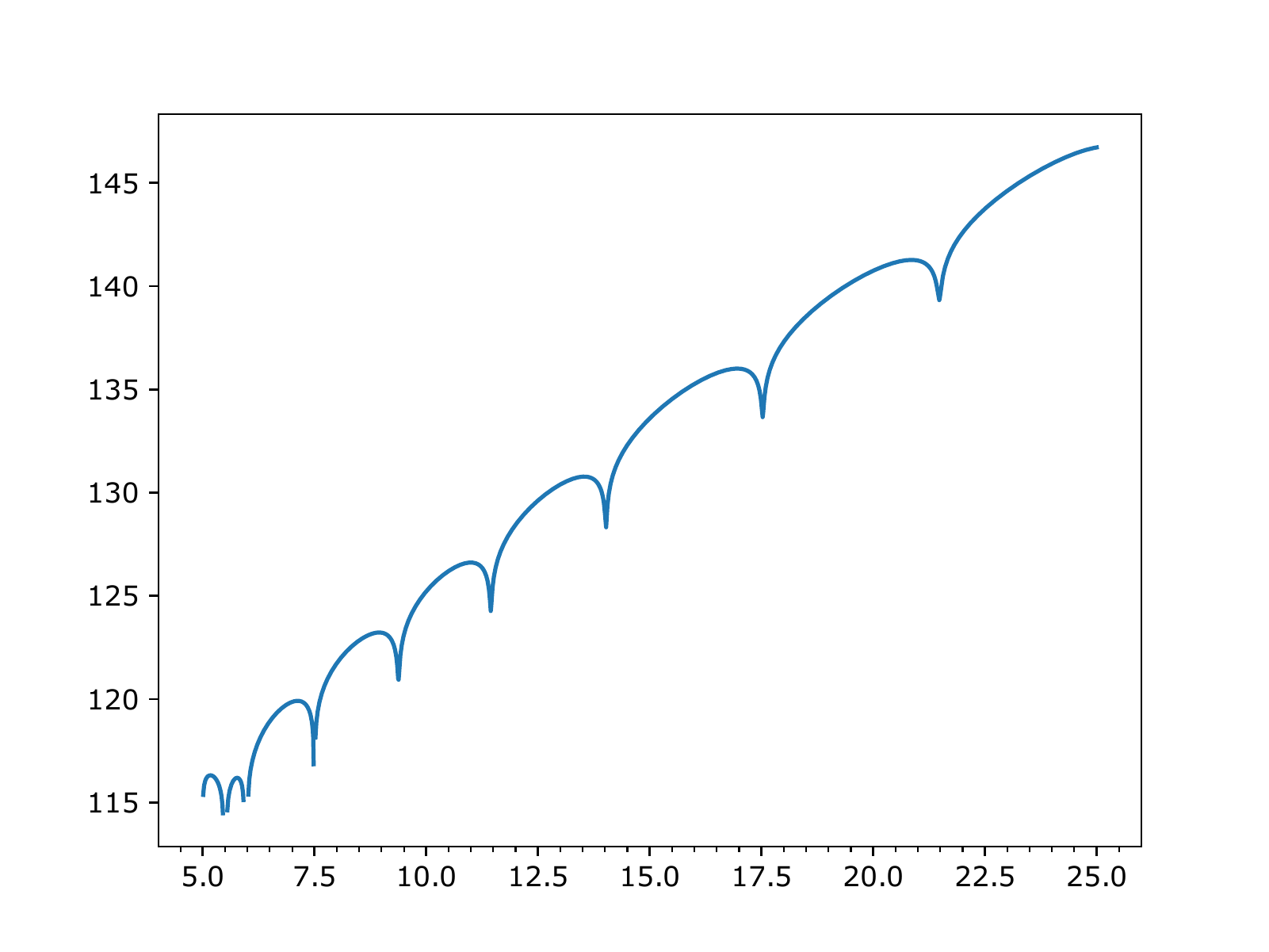}\vspace{-7.5pt}
        \caption{$Y$, $\ell=4$}
    \end{subfigure}
    ~

     \begin{subfigure}[b]{0.27\textwidth}
      \vspace{7.5pt}  \includegraphics[width=\textwidth]{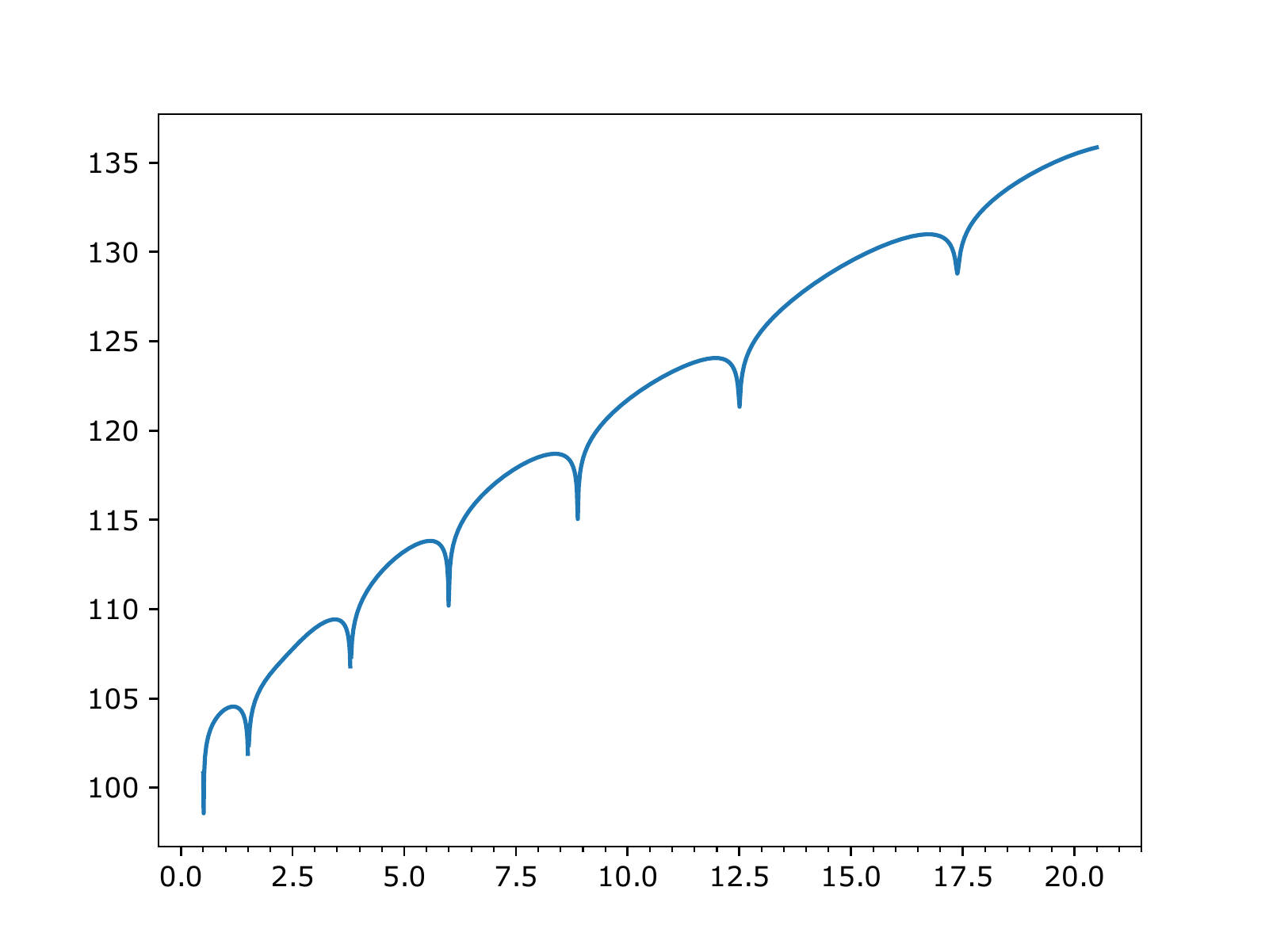}\vspace{-7.5pt}
        \caption{$Z$, $\ell=0$}
    \end{subfigure}
    ~
          \begin{subfigure}[b]{0.27\textwidth}
     \vspace{7.5pt}   \includegraphics[width=\textwidth]{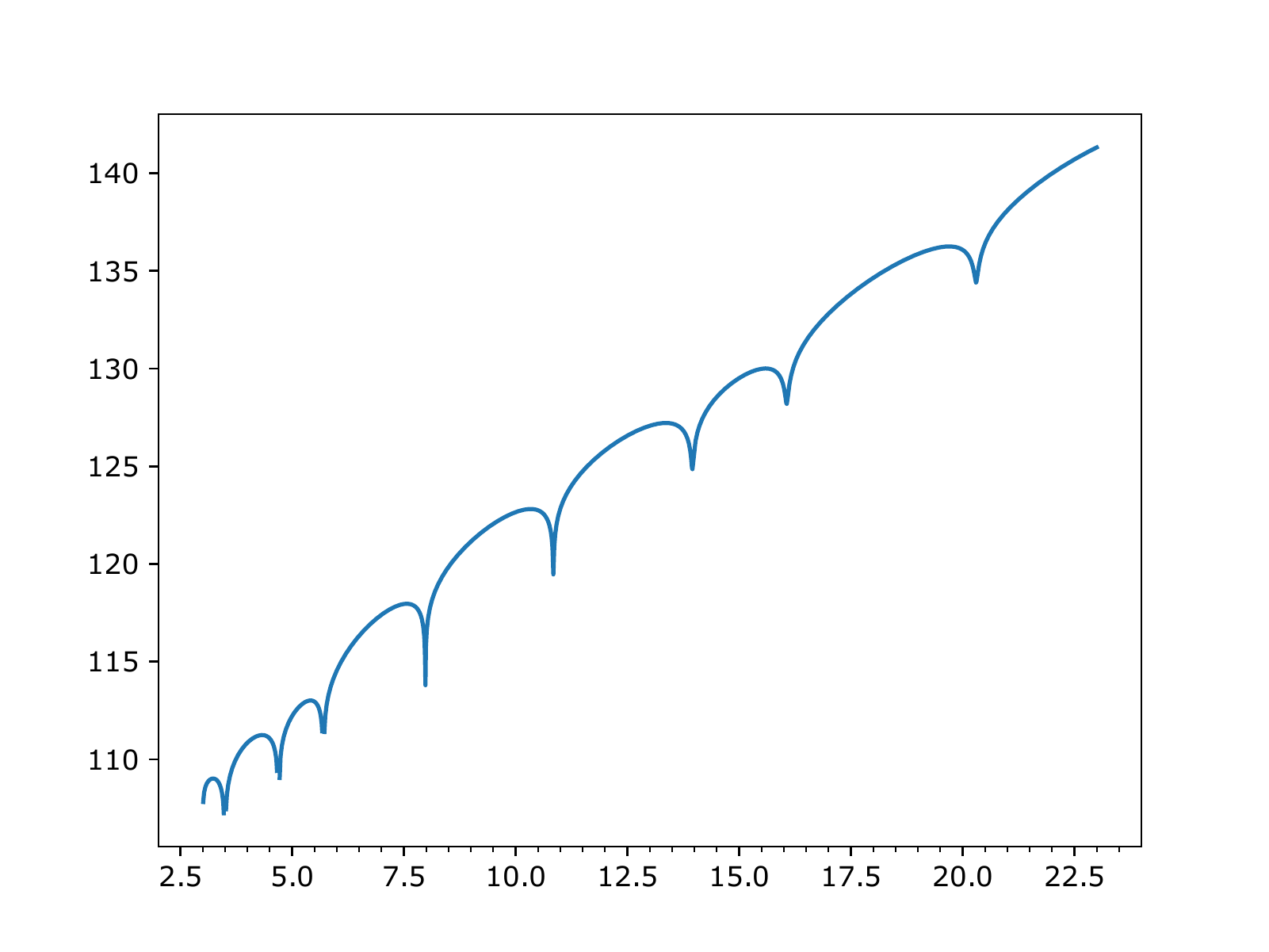}\vspace{-7.5pt}
        \caption{$Z$, $\ell=2$}
    \end{subfigure}
    ~
          \begin{subfigure}[b]{0.27\textwidth}
     \vspace{7.5pt}   \includegraphics[width=\textwidth]{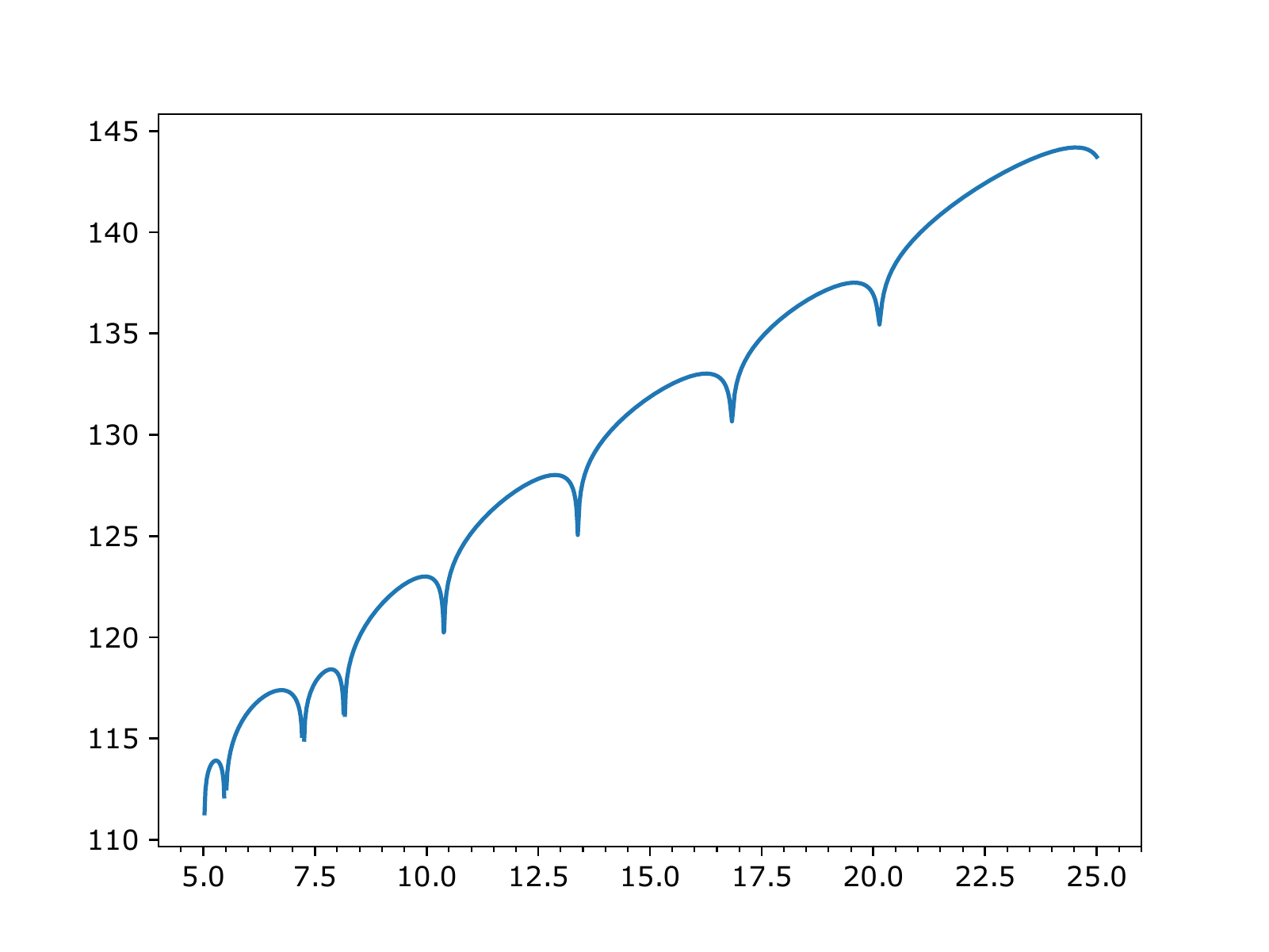}\vspace{-7.5pt}
        \caption{$Z$, $\ell=4$}
    \end{subfigure}
    ~
  \caption{Plots from the extremal functional method for the even
  representations. In the horizontal axis we plot the scaling dimension and
  in the vertical the logarithm of the action of the functional on
  convolved conformal blocks (denoted by $F_{\Delta,\ell}$
  in~\cite{ElShowk:2012hu}).}
  \label{fig:extremal-functional}
\end{figure}

\begin{figure}
  \centering
      \begin{subfigure}[b]{0.27\textwidth}
        \vspace{7.5pt}\includegraphics[width=\textwidth]{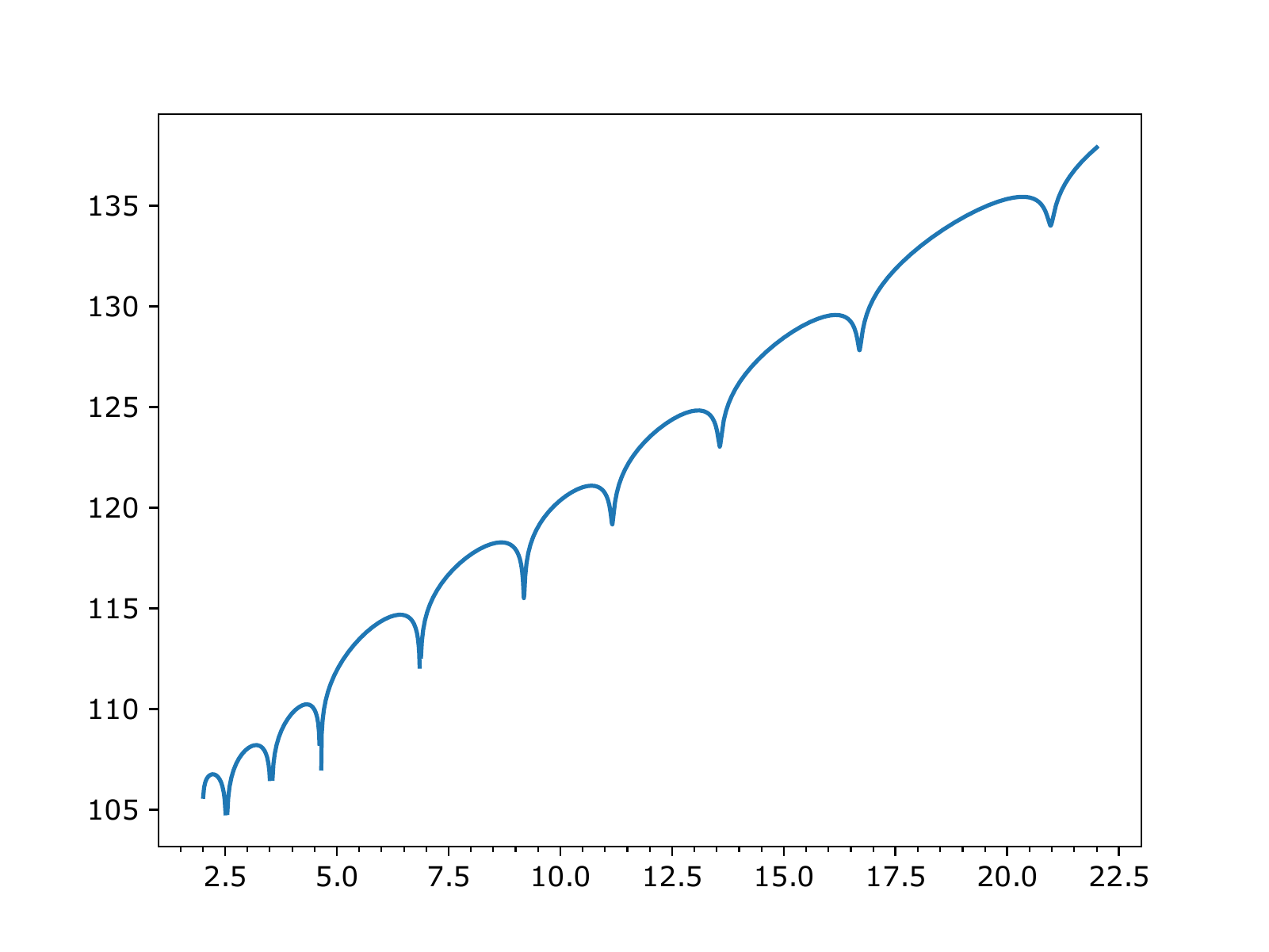}\vspace{-7.5pt}
        \caption{$A$, $\ell=1$}
    \end{subfigure}
    ~
          \begin{subfigure}[b]{0.27\textwidth}
        \vspace{7.5pt}\includegraphics[width=\textwidth]{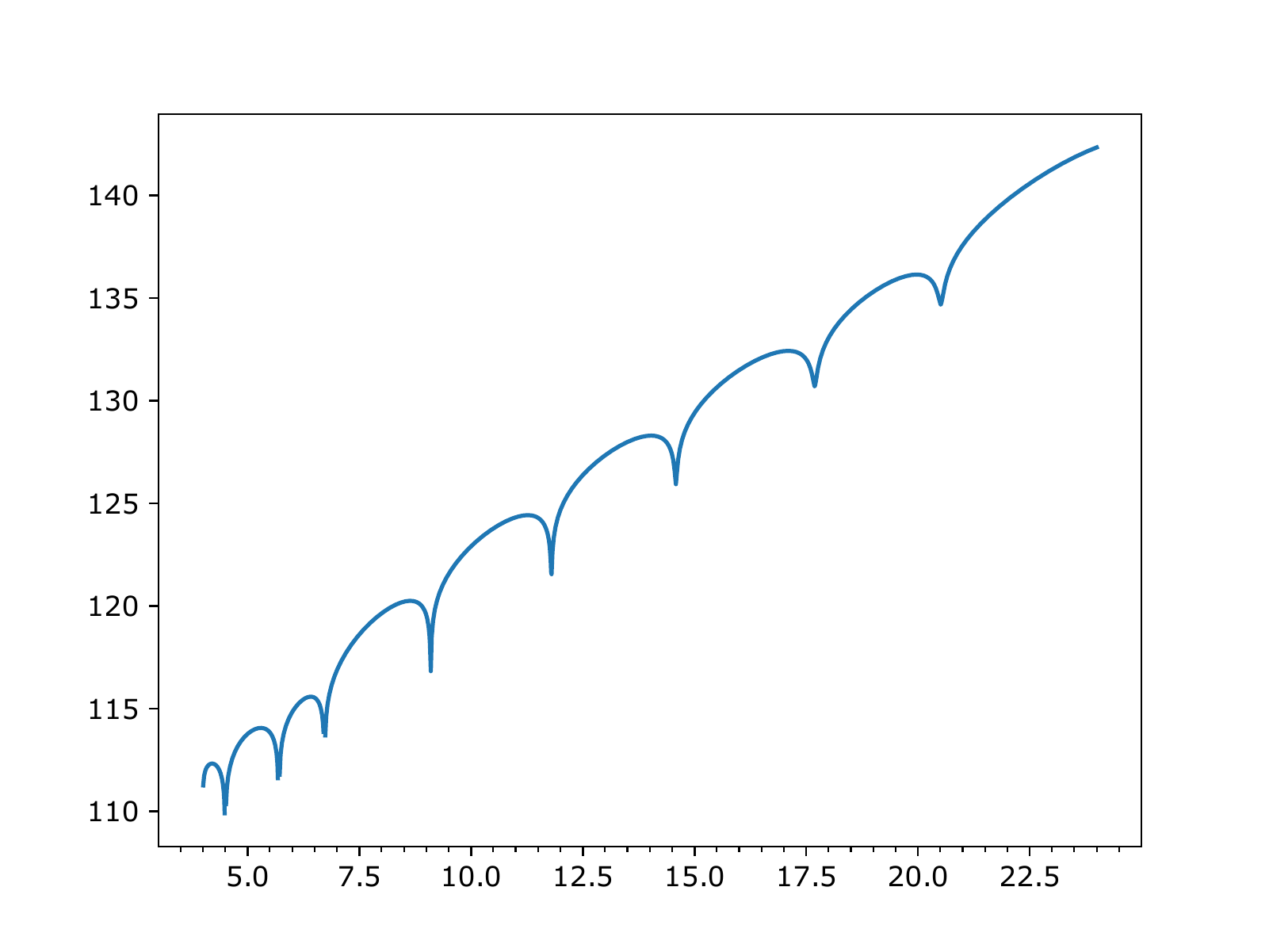}\vspace{-7.5pt}
        \caption{$A$, $\ell=3$}
    \end{subfigure}
    ~
              \begin{subfigure}[b]{0.27\textwidth}
       \vspace{7.5pt} \includegraphics[width=\textwidth]{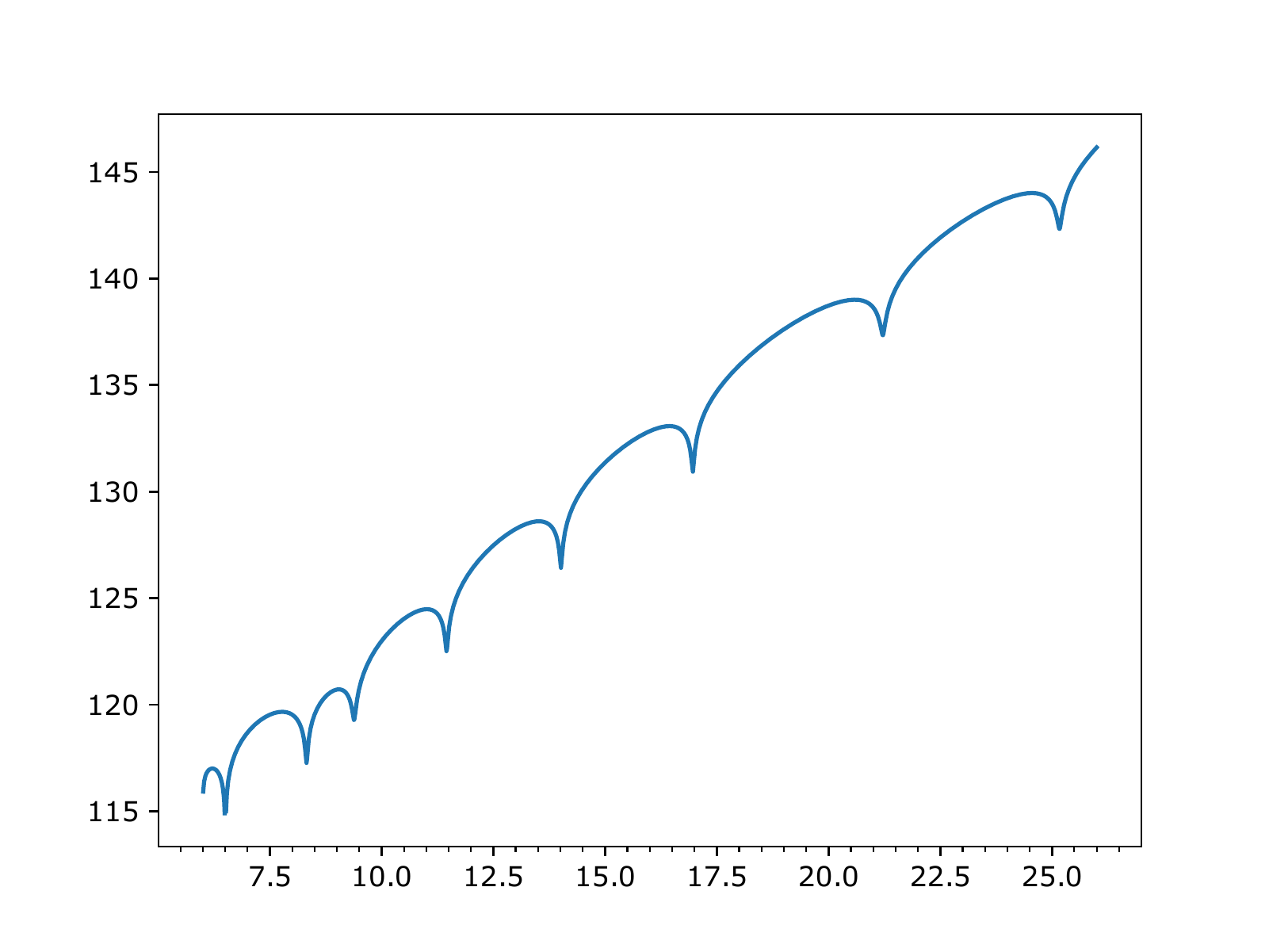}\vspace{-7.5pt}
        \caption{$A$, $\ell=5$}
    \end{subfigure}
    ~

          \begin{subfigure}[b]{0.27\textwidth}
       \vspace{7.5pt} \includegraphics[width=\textwidth]{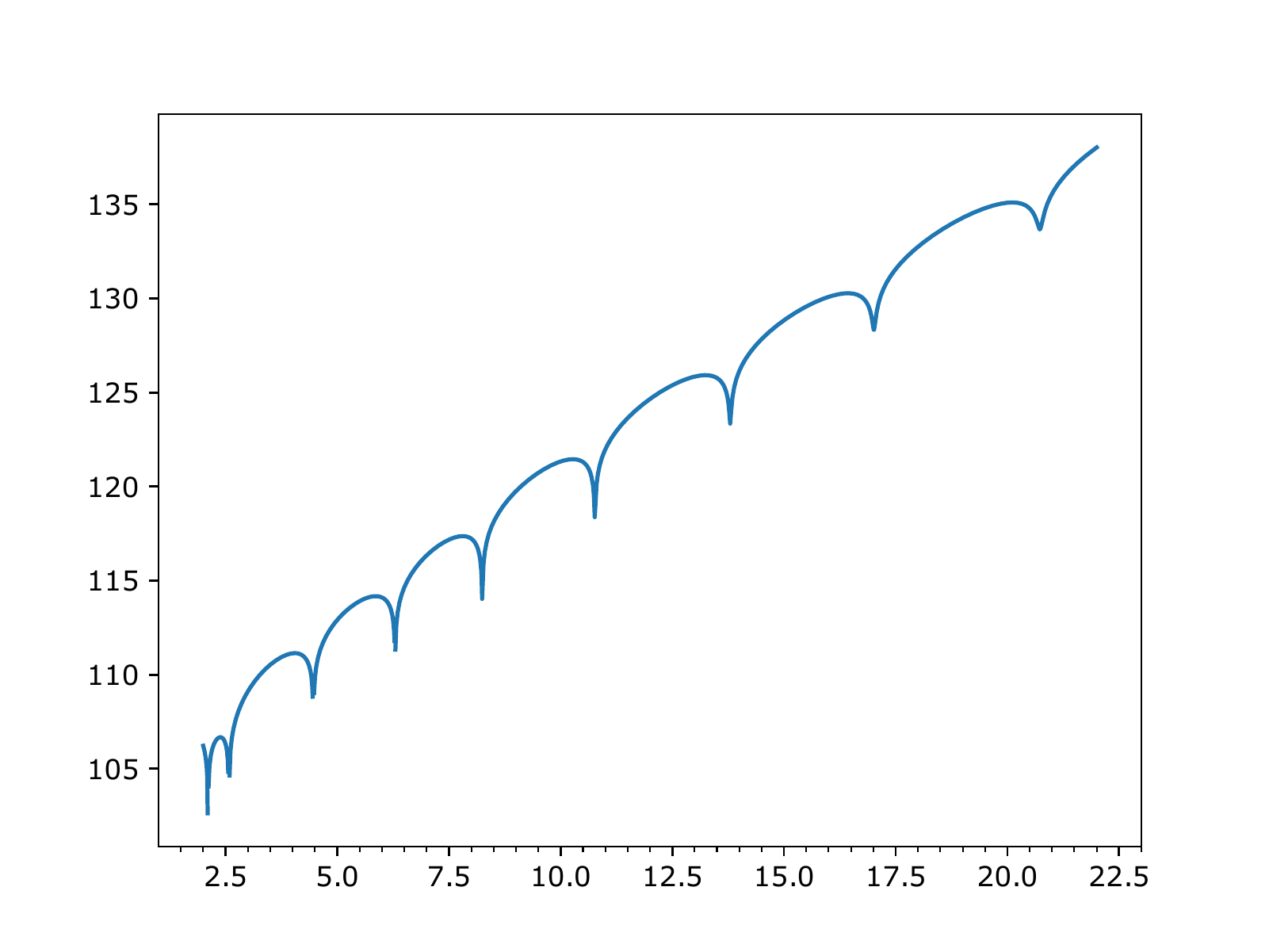}\vspace{-7.5pt}
        \caption{$B$, $\ell=1$}
    \end{subfigure}
    ~
          \begin{subfigure}[b]{0.27\textwidth}
       \vspace{7.5pt} \includegraphics[width=\textwidth]{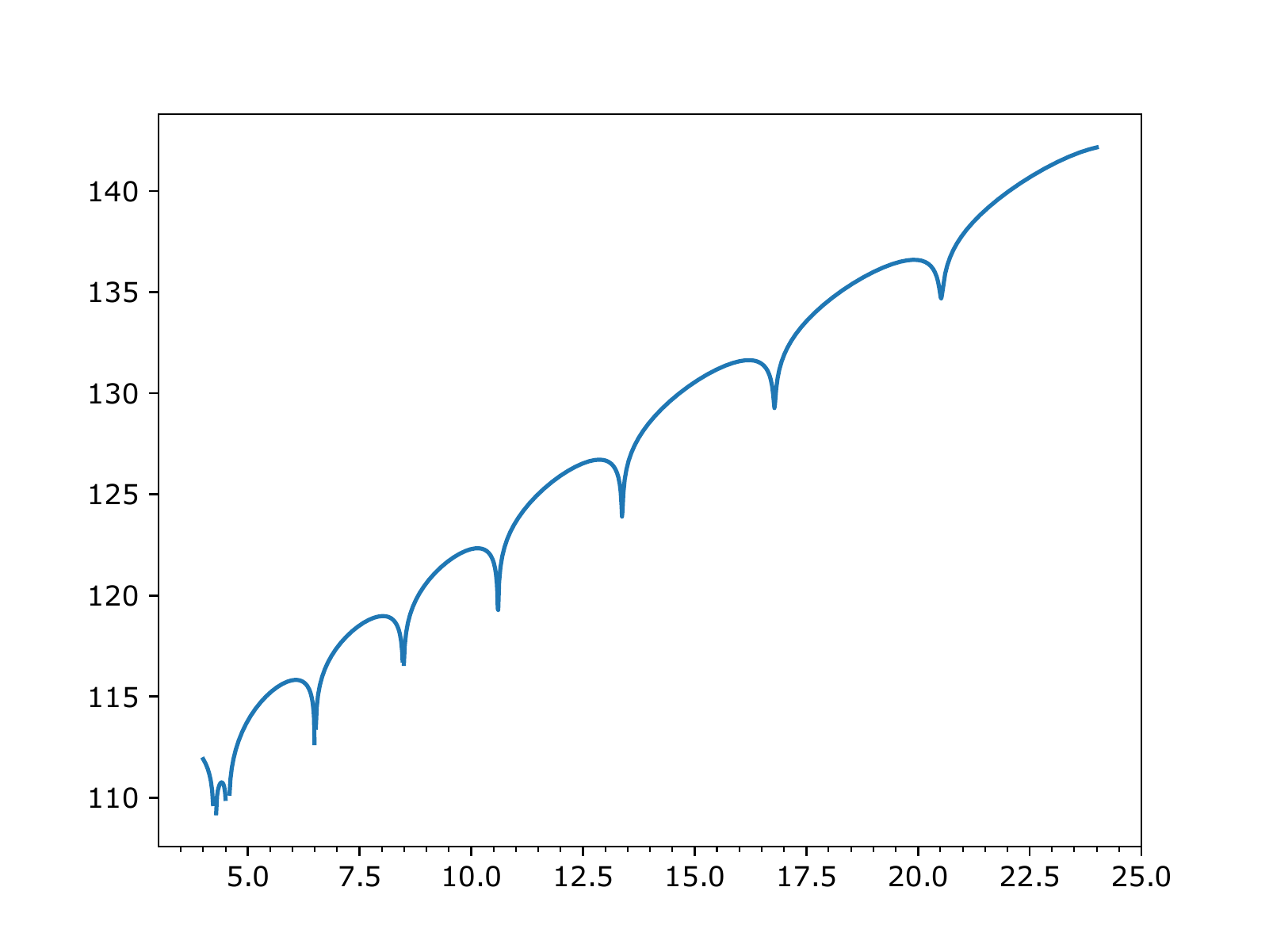}\vspace{-7.5pt}
        \caption{$B$, $\ell=3$}
    \end{subfigure}
    ~
              \begin{subfigure}[b]{0.27\textwidth}
       \vspace{7.5pt} \includegraphics[width=\textwidth]{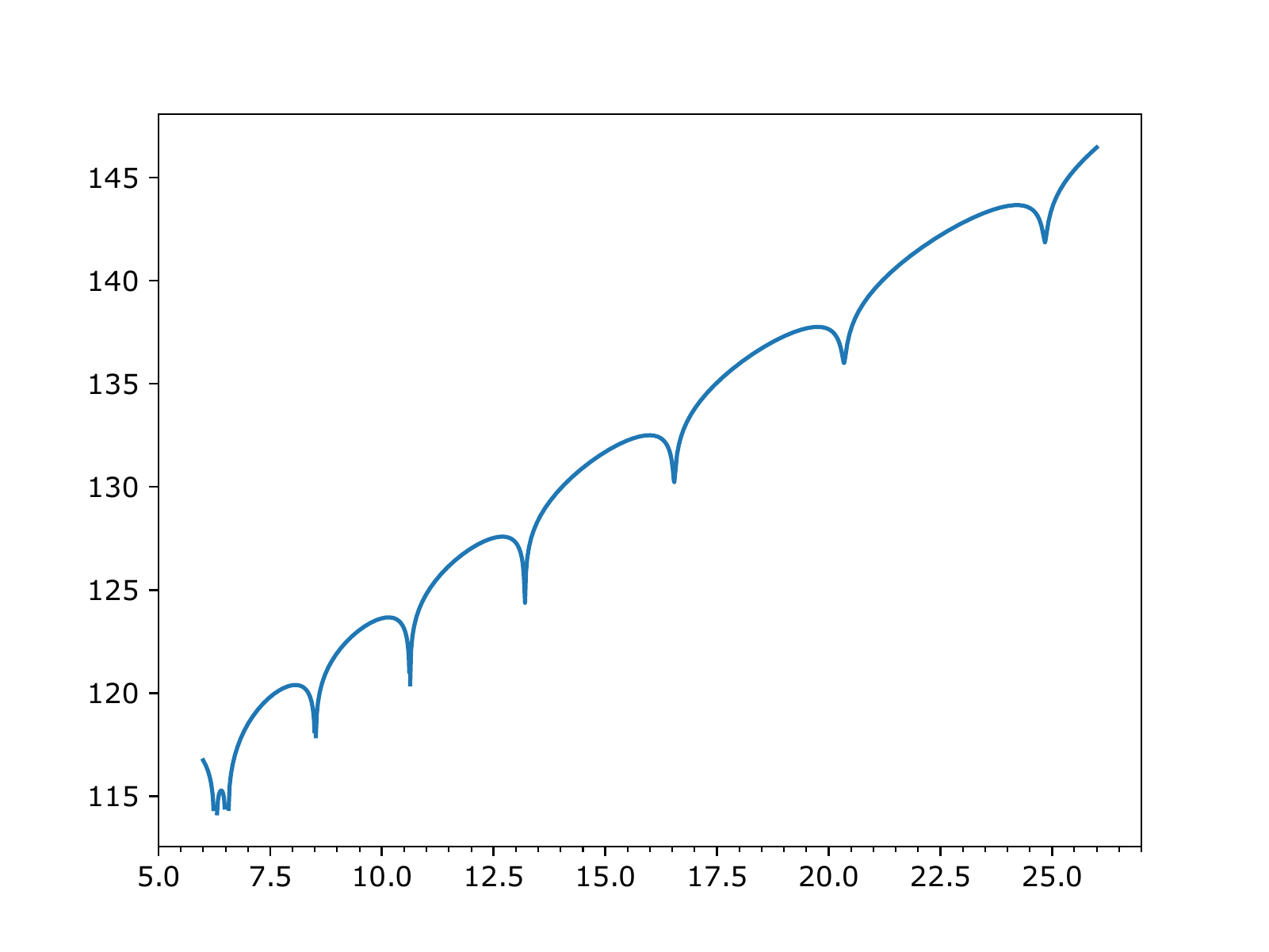}\vspace{-7.5pt}
        \caption{$B$, $\ell=5$}
    \end{subfigure}
    ~
  \caption{Plots from the extremal functional method for the odd
  representations.}
  \label{fig:extremal-functional-odd}
\end{figure}

Subsequently, we focused on the values $(m,n)=(100,2)$, and extracted the
spectrum using the extremal functional method \cite{ElShowk:2012hu}; see
Fig.~\ref{fig:extremal-functional} and
Fig.~\ref{fig:extremal-functional-odd}.  We present results for
$\Delta_\phi=0.75$, for which we used \texttt{PyCFTBoot}
\cite{Behan:2016dtz} with parameters $\texttt{n\_max=13}$,
$\texttt{m\_max=10}$, $\texttt{k\_max=50}$ and $\texttt{l\_max=40}$.
From Fig.~\ref{fig:position-100-2} we we estimate the scalar operator dimensions, for $(m,n)=(100,2)$, to
\begin{align}\label{eq:deltaphi2}
\Delta_\phi&=0.75(1)\,,
\\\label{eq:deltaX2}
\Delta_X&=6.1(4)\,.
\end{align}
From Figs.~\ref{fig:extremal-functional}
and~\ref{fig:extremal-functional-odd} we then read off the dimensions of the leading scalar operators to
\begin{align}
\label{eq:deltaS2}
\Delta_S&\approx1.35\,,
\\\label{eq:deltaY2}
\Delta_Y&\approx0.8\,,
\\\label{eq:deltaZ2}
\Delta_Z&\approx0.6\,.
\end{align}
It is interesting to note that these results have $\Delta_Z<\Delta_\phi$.
The small values for $\Delta_Y$ and $\Delta_Z$ suggest that a mixed
correlator bootstrap involving the operators $Y$ and/or $Z$ may give
results that are quite constraining.

\subsec{Nonperturbative aspects of the large \texorpdfstring{$m$}{m} theory}

The results from our numerical bootstrap show that the second kink
continues to exist for all values of $m\geq2$, indicating the existence of
a corresponding CFT, in the sense of a set of conformal primary operators
with scaling dimensions and OPE coefficients consistent with unitarity and
crossing. We have only considered the constraints from the
$\langle\phi\phi\phi\phi\rangle$ correlator, and numerical studies using a
multi-correlator approach will either give further constraints on the
candidate CFT or disprove its existence.

The motivation for our work was to see if the candidate theory approaches a
simplifying limit as $m\to\infty$, for instance a perturbation from a theory of generalized free fields. If this were the case, it could potentially be
studied using perturbative methods just like we did for the first kink in
section~\ref{sec:kink1}. Our results show that this is not the case,
meaning that the candidate CFT remains non-perturbative, or ``strongly
coupled'' for all values of $m$. It will therefore be challenging to corroborate the existence of this CFT.

There is, however, one test that any conformal field theory must pass, be it perturbative or not, namely consistency with the predictions from the lightcone bootstrap \cite{Fitzpatrick:2012yx,Komargodski:2012ek}. These papers proved the \emph{twist additivity}, stating that any conformal field theory containing operators $\O_1$ and $\O_2$, must also contain an infinite family of spinning operators $[\O_1,\O_2]_{0,\ell}$ \footnote{The interacting theory also contains subleading twist families $[\O_1,\O_2]_{k,\ell}$ with twists approaching $\tau_1+\tau_2+2k$.} with twists (recall that $\tau=\Delta-\ell$)
\begin{equation}
  \lim_{\ell\to\infty}\tau_{\ell}=\tau_\infty= \tau_1+\tau_2\,.
\end{equation}
This statement means that a given theory will contain many different accumulation
points $\tau_\infty$, see \cite{Kehrein:1995ia} for observations in the
$\varepsilon$ expansion of the $O(N)$ model predating the lightcone bootstrap, however not all these values
may be visible in a given correlator. In fact, the only accumulation point
guaranteed to exist in the direct channel of the
$\langle\O_1\O_2\O_2\O_1\rangle$ is $\tau_1+\tau_2$, but other values are
not excluded. The conclusion is that we expect that in our candidate CFT
the value
\begin{equation}\label{eq:twodeltaphi}
\tau_\infty=2\Delta_\phi=1.50(2)
\end{equation}
is a twist accumulation point, and operators in the twist family
corresponding to this value should be visible in our numerical bootstrap
study.

In a weakly coupled theory, where $\Delta_\phi$ is close to the scalar
unitarity bound $\frac{d-2}2$, the twists of the double-twist operators
$[\phi,\phi]_{0,\ell}$ are also close to the spinning unitarity bound
$d-2$, c.f.\ \eqref{eq:UB}. In such a theory, the double-twist operators
have leading twist, and will include the stress-energy tensor at $\ell=2$.

In the generic case, where $\Delta_\phi-\frac{d-2}2$ is finite, there are
three possibilities:
\begin{itemize}
\item The double-twist operators $[\phi,\phi]_{0,\ell}$ remain the
operators at leading twist for each spin, which means that they acquire
``large'' anomalous dimensions $\gamma_\ell=\tau_\ell-\tau_\infty$ in order
to accommodate the stress-energy tensor at $\ell=2$. This is the situation in 3D
critical $O(N)$ models at finite $N$, and indeed also in the perturbative
fixed point of section~\ref{sec:kink1} in this paper.\footnote{Note that in
these theories, the anomalous dimensions never become particularly large;
the largest values is attained in the 3D $O(2)$ CFT with
$\tau_\infty-\tau_T=0.0383$.}
\item The stress-energy tensor belongs to an additional twist family below the double-twist operators. This  case happens for instance when $\phi$ is a composite operator in weakly coupled theories. An example is correlators of gauge-invariant operators in weakly coupled $\mathcal N=4$ SYM, where the double-twist family consists of double-trace operators and the additional leading twist family consists of the single-trace weakly broken currents.
\item The stress-energy tensor and other conserved currents are isolated
operators not belonging to any twist family. This happens in $\mathcal N=4$
in the strong coupling expansion, where the lower limit of analyticity in spin is
shifted upwards \cite{Caron-Huot:2017vep}, but is not expected for a
\emph{bona fide} CFT with usual Regge behavior.
\end{itemize}
We now study the plots in Figs.~\ref{fig:extremal-functional} and
\ref{fig:extremal-functional-odd} to see if they are consistent with any of
the mentioned scenarios. Since the position of the kink could not be
precisely determined, we do not expect these plots to give very precise
values for the operator dimensions. We can, however, look at the
qualitative behaviour of the lowest dimension operators.

\begin{figure}[ht]
  \centering
{
\setlength{\unitlength}{0.1\textwidth}
  \begin{picture}(10,5.4)
\put(0,0){\begin{subfigure}[b]{0.48\textwidth}

       \includegraphics[width=1\textwidth]{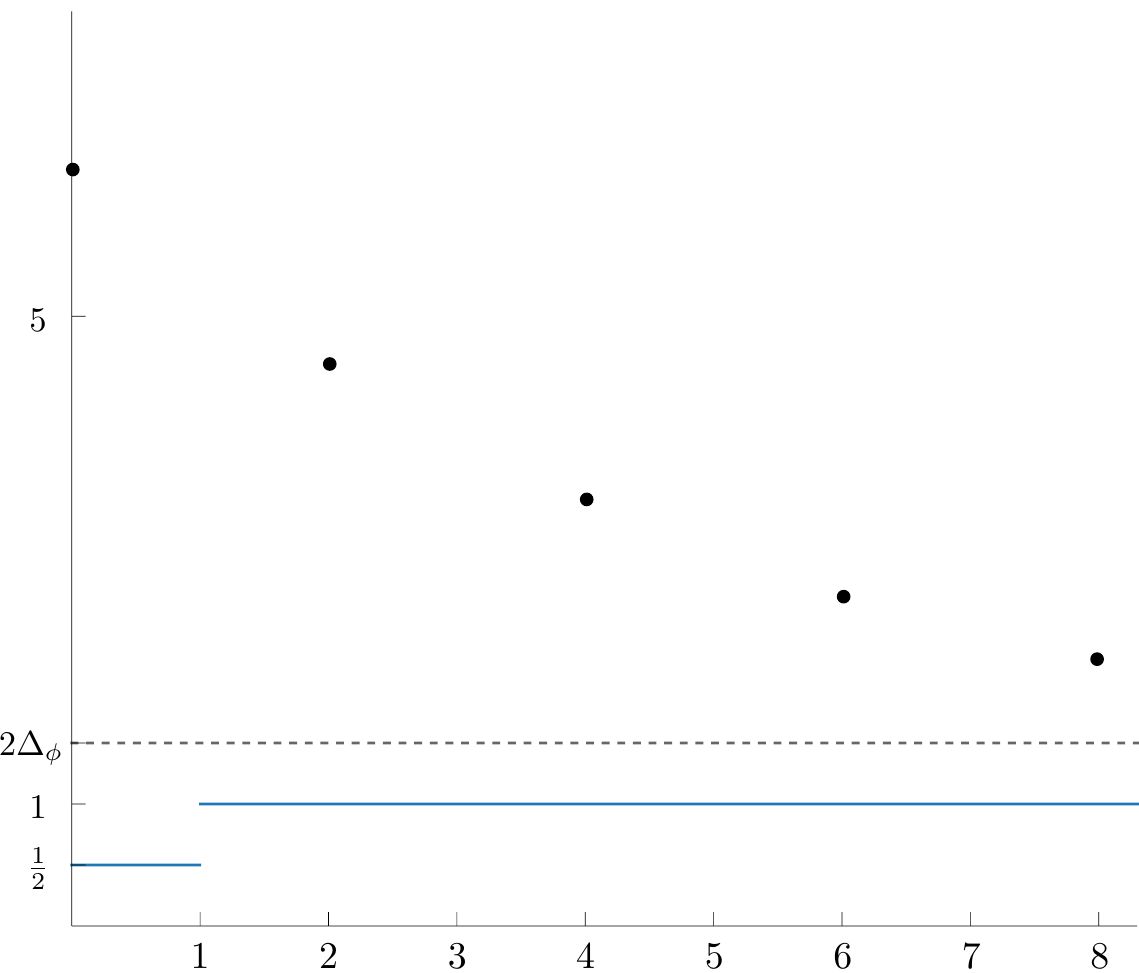}
        \caption{Twist spectrum in $X$.}\label{subfig:tX}
    \end{subfigure}}
\put(5,3.6){\begin{subfigure}[b]{0.48\textwidth}
        \includegraphics[width=1\textwidth]{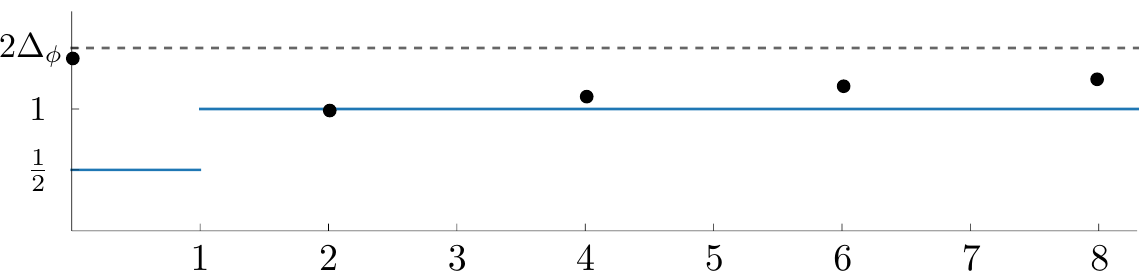}
        \caption{Twist spectrum in $S$.}\label{subfig:tS}
    \end{subfigure}}
\put(5,1.8){\begin{subfigure}[b]{0.48\textwidth}
        \includegraphics[width=1\textwidth]{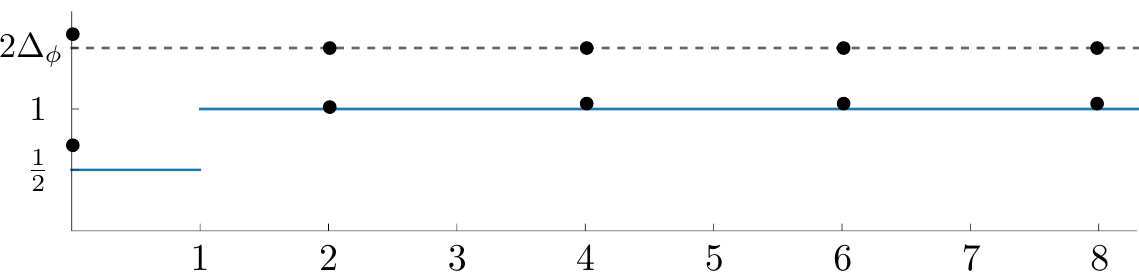}
        \caption{Twist spectra in $Y$ and $Z$.}\label{subfig:tYZ}
    \end{subfigure}}
\put(5,0){\begin{subfigure}[b]{0.48\textwidth}
       \includegraphics[width=1\textwidth]{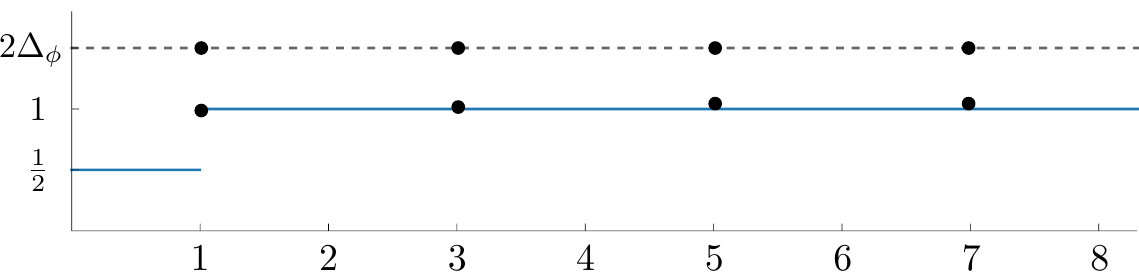}
        \caption{Twist spectra in $A$ and $B$.}\label{subfig:tAB}
    \end{subfigure}}
\end{picture}
  }

  \caption{Cartoons for the hypothetical twist families in the various representations showing spin on the horizontal axis and twist on the vertical axis. This picture is based on the plots in Figs~\ref{fig:extremal-functional} and~\ref{fig:extremal-functional-odd} using the extremal functional method, and needs to be confirmed or disproved by other methods.
  The unitarity bound \eqref{eq:UB} is shown in blue and the value $2\Delta_\phi=1.5$
  is shown in dashed gray. The complete spectrum is expected to contain more operators than those in the Regge trajectories shown. Only $S_2=T$ and $A_1=J$ have twists exactly on the unitarity bound.}
  \label{fig:cartoon}
\end{figure}

In the $S$ and $X$ representations, the spectrum plots are comparatively
sparse. There the lowest dimension operators at each spin are consistent
with the first scenario, with large anomalous dimensions. In the singlet
representation, these anomalous dimensions are negative, which is
consistent with Nachtmann's theorem (convexity) and with the stress-energy tensor
appearing at $\ell=2$. In the $X$ representation, the anomalous dimensions
are positive, and it seems like the family can be extended to spin zero to
include the $X$ operator.\footnote{By studying also the spectrum plots at
spin $6$ and $8$ (not displayed), we note that the twists of the leading
operator continues to decrease according to the behavior displayed in
Fig.~\ref{fig:cartoon}. Note however the feature at $\Delta=5$ in
Fig.~\ref{subfig:X4}, a similar feature is noted at spin $6$ (but not at
spin $8$) and is likely to be a spurious zero.} On the contrary, the $Y$,
$Z$, $A$, and $B$ representations show a comparably dense spectrum, which
is consistent with the second scenario of a twist family below the
double-twist operators.  In Fig.~\ref{fig:cartoon} we display a cartoon
that summarizes these observations; however, further numerical study will be
needed to confirm or disprove this picture.

While it is encouraging that the spectrum plots are not inconsistent with
the constraints from the lightcone bootstrap, we would like to address some
issues that complicate the picture. As already mentioned, the uncertainty
in the position of the kink induces uncertainty in the spectrum plots.
Moreover, the plots should not be interpreted as displaying the full
spectra in the respective representations since they are only showing
operators with a non-negligible contribution to the $\phi$ four-point
function. The complete spectrum of the theory is more dense than our
cartoons of Fig.~\ref{fig:cartoon} indicate. For instance, the singlet
representation is expected to contain accumulation points $2\tau_\O$ for
all operators $\O$ in the theory and in our case this would include an
accumulation point $2\Delta_Z=1.2$, which is below the value $2\Delta_\phi$
in \eqref{eq:twodeltaphi}. A numerical bootstrap study of mixed correlators may reveal more operators. In
\cite{Simmons-Duffin:2016wlq}, an ambitious attempt was made to determine
the operator spectrum in the 3D Ising CFT, using a system of mixed
correlators constructed out of $\sigma$ and $\epsilon$, the smallest
dimension $\mathbb Z_2$ odd and even operators. Based on
\cite{Komargodski:2016auf}, the extremal functional was then applied to a
sample of points on the boundary of the allowed region (in this case an
island), and only stable operator dimensions were deemed to be candidates
for primary operators in the spectrum. A similar computation was performed
in \cite{Liu:2020tpf} for the 3D $O(2)$ CFT.  The method appears to give
somewhat reliable predictions for the spectrum at low operator dimensions,
but misses higher-spin operators asymptoting to double-twist dimensions of
operators not included in the system of correlators studied.\footnote{For
instance, results of \cite{Simmons-Duffin:2016wlq} clearly identify the
accumulation points $2\Delta_\sigma=1.036$, $2\Delta_\epsilon=2.825$ and
$2\Delta_\sigma+2=3.036$, but misses the intermediate values $2\tau_T=2$
and $\tau_T+\Delta_\epsilon=2.413$. Likewise in the $O(2)$ CFT, not all
expected operators in the charge $4$ sector were found numerically in
\cite{Liu:2020tpf}.} It would be desirable to perform a similar study of the theory at hand.

\newsec{Discussion}
In this work we studied CFTs with global symmetry $\MNmath_{m,n} =
O(m)^n\rtimes S_n$ in $d=3$ dimensions, with the motivation of evaluating
further the potential existence of two distinct fixed points in such
theories, as was recently suggested in~\cite{Stergiou:2019dcv} for the
$\MNmath_{2,2}$ and $\MNmath_{2,3}$ theories. For theories with $n=2$ we
found evidence supporting this conclusion by considering various values of
$m$ and observing two distinct kinks in bootstrap bounds, even for large
$m$. For $n>2$, our bootstrap bounds did not include a clear second kink in
the expected region for large $n$, although such a kink does exist in the
$\MN_{2,3}$ theory (see \cite[Fig.~4]{Stergiou:2019dcv}). Our results
suggest that there exists a critical line $n_c(m)$ below which there are
two distinct \MN CFTs.

The second kink we examined appears to not correspond to a known theory
obtained within the standard Wilson--Fisher paradigm. Looking at the
operator spectrum at this kink, we verified that it satisfies general
expectations derived from the Nachtmann theorem~\cite{Nachtmann:1973mr} and
the lightcone bootstrap \cite{Fitzpatrick:2012yx,Komargodski:2012ek}. If
this kink is due to a corresponding full-fledged CFT, this would indicate
that the Wilson--Fisher paradigm is incomplete, i.e.\ that there exist
fixed points in $d=3$ that cannot be obtained from continuations of fixed
points in $d=4-\varepsilon$.  Recently, other bootstrap works have reported
kinks that do not appear to be of Wilson--Fisher type \cite{Li:2020bnb,
He:2020azu, Reehorst:2020phk}, but the qualitative features of these kinks
differ from ours and they may be of different origin. Interestingly, the
picture emerging from our spectrum analysis at large $m$ shows some
similarities with results in large $N$ $O(N)^3$ bosonic tensor models
\cite{Giombi:2017dtl}.\footnote{In these models, $\Delta_\phi=d/4$, and the
bilinear operators in some representations, including those containing
conserved currents, acquire large anomalous dimensions, similar to
Fig.~\ref{subfig:tX} and~\ref{subfig:tS}; see
\cite{Popov:2019nja,Lettera:2020uay} for recent work in the supersymmetric
case.} It would be interesting to further investigate the consistency of
the spectrum of the second kink, using a numerical application of the
Lorentzian inversion formula as has been done for the Ising and $O(2)$ CFTs
\cite{Liu:2020tpf,Caron-Huot:2020ouj}; see also
\cite{Simmons-Duffin:2016wlq}. This would require a more precise
determination of the spectrum and estimation of OPE coefficients.

With regards to the experimentally accessible $\MNmath_{2,2}$ case, out of
the five scenarios of \cite[Sec.\ III.B.3]{Delamotte:2003Dw}, all of our
bootstrap results favor ``Scenario II'', with two fixed points and
therefore second-order phase transitions in both groups of systems.
However, we stress that the values of the exponent $\beta$ obtained in
experiments are in mild tension with the bootstrap ones for the first kink,
see Tab.~\ref{table:comparison-kink1}.  Related to this issue is perhaps
the fact that experimental and Monte Carlo results are inconsistent with
unitarity, which predicts that $2\beta-\nu\geq0$. Of course, by
construction the numerical bootstrap results are consistent with unitarity.
Given the relatively good agreement of $\nu$ between bootstrap,
experimental and Monte Carlo results, we conclude that there is need for a
more accurate determination of $\beta$ with experiments and Monte Carlo
simulations, controlling systematic errors.

It would be desirable to compute results at large $m$ using conventional
diagrammatic techniques similar to the large $N$ expansion of the critical
$O(N)$ model. There, the leading scalar operator dimensions have been
computed to orders $N^{-3}$ ($\Delta_\phi$ \cite{Vasiliev:1982dc}) and
$N^{-2}$ ($\Delta_S$ \cite{Okabe:1978mp,Vasiliev:1981dg} and $\Delta_T$
\cite{Gracey:2002qa}), and results for the next-to-leading scalar singlet
at order $N^{-2}$ also exist~\cite{Broadhurst:1996ur}. A difference with
the $O(N)$ model is that the auxiliary field is now $X$, which is not a
singlet under the global symmetry.

Additionally, it would be informative to perform a numerical bootstrap in
intermediate dimensions $3<d<4$, like that in \cite{El-Showk:2013nia,
Cappelli:2018vir, Stergiou:2018gjj}, in order to examine the behavior of
the kinks as we approach $d=4$. This will allow us to make better contact
with the $\varepsilon$ expansion for the first kink, as well as determine
if the second kink persists closer to $d=4$. We note here that theories
defined for non-integer values of $d$ are expected to be non-unitary
\cite{Hogervorst:2015akt}, but this is not expected to cause problems when
bounding scaling dimensions of low-lying operators. Mixed correlator
bootstrap studies in $d=3$ and $3<d<4$ could also yield crucial pieces of
information that would allow us to further characterize the kinks.
Finally, it would be instructive to apply alternative methods, such as
non-perturbative RG, to \MN theories; see \cite{Chlebicki:2020pvo} for
successful recent work away from unitarity in $O(N)$ models.

\ack{
We would like to thank J.\ Gracey and A.\ Vichi for helpful discussions and
comments on the manuscript. The project has received partial funding from
the European Research Council (ERC) under the European Union's Horizon 2020
research and innovation programme (grant agreement no. 758903). This
research used resources provided by the Los Alamos National Laboratory
Institutional Computing Program, which is supported by the U.S.\ Department
of Energy National Nuclear Security Administration under Contract No.\
89233218CNA000001.  Research presented in this article was supported by the
Laboratory Directed Research and Development program of Los Alamos National
Laboratory under project number 20180709PRD1.
}

\begin{appendices}

\newsec{Large \texorpdfstring{$\boldsymbol{m}$}{m} results for general \texorpdfstring{$\boldsymbol{d}$}{d}}\label{sec:app}

In this appendix we present the results of section~\ref{sec:LSPT} for
general spacetime dimension $d=2\mu$. The scalar operator dimensions are
given by
\begin{align}\label{eq:Ddeltaphi}
  \Delta_\phi&=\mu-1+\frac{n-1}n\frac{\eta_1}{m}+O(m^{-2})\,,
\\\label{eq:DdeltaS}
\Delta_S&=2(\mu-1)+\frac{4 (\mu -1) (2 \mu -1) (n-1)}{(2-\mu )
n}\frac{\eta_1}{m}+O(m^{-2})\,,
\\\label{eq:DdeltaX}
\Delta_X&=2-4\frac{ (\mu -1) (2 \mu -1) n-4 \mu ^2+6 \mu -1}{(2-\mu )
n}\frac{\eta_1}{m}+O(m^{-2})\,,
\\\label{eq:DdeltaY}
\Delta_Y&=2(\mu-1)+\frac{4 (n-1)}{(2-\mu ) n}\frac{\eta_1}{m}+O(m^{-2})\,,
\\\label{eq:DdeltaZ}
\Delta_Z&=2(\mu-1)+2\frac{(2- \mu  )n-2}{(2-\mu )
n}\frac{\eta_1}{m}+O(m^{-2})\,,
\end{align}
where $
\eta_1=\frac{(\mu -2) \Gamma (2 \mu -1)}{\Gamma (1-\mu ) \Gamma (\mu )^2 \Gamma (\mu +1)}$; the spinning operator dimensions by
\begin{align}
\Delta_{S_\ell}&=\ell+2\Delta_\phi-\frac{2 \mu  (n-1)}{J^2n}
\left(\mu-1+\frac{\Gamma (\ell+1) \Gamma (2 \mu -1)}{\Gamma (\ell+2 \mu
-3)}\right)\frac{\eta_1}{m}+O(m^{-2})\,,
\\
\Delta_{X_\ell}&=\ell+2\Delta_\phi-\frac{2 \mu}{J^2n}\left((\mu -1)
(n-1)+\frac{(n-2) \Gamma (\ell+1) \Gamma (2 \mu -1)}{\Gamma (\ell+2 \mu
-3)}\right)\frac{\eta_1}{m}+O(m^{-2})\,,
\\
\Delta_{Y_\ell}=\Delta_{A_\ell}&=\ell+2\Delta_\phi-\frac{2 (\mu -1) \mu
(n-1)}{J^2n}\frac{\eta_1}{m}+O(m^{-2})\,,
\\
\Delta_{Z_\ell}=\Delta_{B_\ell}&=\ell+2\Delta_\phi+\frac{2 \mu(\mu -1)
}{J^2n}\frac{\eta_1}{m}+O(m^{-2})\,,
\end{align}
where $J^2=(\mu-1+\ell)(\mu-2+\ell)$ and we have substituted the value for $a_X=\frac{ (n-1)(\mu -1) \mu }{ n(\mu -2)^2}\eta_1$; and the central charge corrections by
\begin{align}
\frac{C_T}{C_{T,\mathrm{free}}}&=1-\frac{n-1}{(2-\mu) \mu  (\mu +1) n}\left(
2 \mu(2-\mu)[ \pi\cot (\pi  \mu ) +  S_1(2 \mu -2)]-\mu ^2+2 \mu+4
\right)\frac{\eta_1}{m}+O(m^{-2})\,,
\\
\frac{C_J}{C_{J,\mathrm{free}}}&=1-\frac{2 (2 \mu -1) (n-1)}{\mu(\mu -1)
n}\frac{\eta_1}m+O(m^{-2})\,,
\end{align}
where $S_1(x)$ denotes the standard analytic continuation of the harmonic numbers away from integer arguments. For $\mu=3/2$ the expressions here reduce to those computed in section~\ref{sec:LSPT}, and for $\mu=2-\varepsilon/2$, the expressions agree with the known results in the $\varepsilon$ expansion.

\end{appendices}

\bibliography{biblio}

\end{document}